\documentclass[aps,twocolumn,nofootinbib,prd,eqsecnum,showpacs,showkeys,preprintnumbers]{revtex4-1}
\usepackage[caption=false]{subfig}
\usepackage{graphicx}
\usepackage{amsmath}
\usepackage{amsfonts}
\usepackage{amssymb}
\usepackage{color}
\usepackage{bm}
\usepackage{mathrsfs}
\usepackage{epstopdf}
\usepackage{url}
\usepackage{footnote}
\usepackage{textcomp}
\usepackage{ulem}
\usepackage{esint}
\usepackage[unicode=true, pdfusetitle,
 bookmarks=true,bookmarksnumbered=false,bookmarksopen=false,
 breaklinks=false,pdfborder={0 0 1},backref=false,colorlinks=false]{hyperref}
\usepackage{multirow}
\usepackage{pifont}

\begin{document}
\title{ Induced gravity effect on inflationary parameters in an holographic cosmology}

\author{Aatifa Bargach}
\email{a.bargach@ump.ac.ma}
\author{Farida Bargach}
\email{f.bargach@ump.ac.ma}
\author{Ahmed Errahmani}
\email{ahmederrahmani1@yahoo.fr}
\author{Taoufik Ouali}
\email{ouali1962@gmail.com}
\affiliation{Laboratory of Physics of Matter and Radiation, \\
University of Mohammed first, BP 717, Oujda, Morocco}

\date{\today }

\begin{abstract}
We investigate observational constraints on inflationary parameters in the context of an holographic cosmology with an induced gravity correction. We consider two situations where a universe is firstly filled with a scalar field and secondly with a tachyon field. Both cases are investigated in a slow-roll regime. We adopt a quadratic potential and an exponential potential for the scalar and the tachyon inflation respectively. In this regard, the standard background and perturbative parameters characterizing the inflationary era are modified by correction terms. We show a good agreement between theoretical model parameters and Planck2018 observational data for both scalar and tachyon fields.

\end{abstract}
\maketitle
\section{Introduction}

One of the rigorous tasks facing cosmology is to obtain a simple solution to many problems of standard cosmology, such as the  horizon, the flatness and the  monopole problems. Primordial inflation is the most successful paradigm to describe the early universe and to solve the aforementioned issues \cite{inflation1, inflation2}. Even though, the inflationary model provides a natural explanation for the origin of primordial perturbations \cite{inflation3, inflation4}, general relativity breaks down at high enough energies and then it should be modified \cite{gr}. Several models of modification of gravity in which inflation can be realized have been developed \cite{jcap, boer, mariam, Lidsey:2005nt, brane, Kofinas:2001es, Deffayet:2000uy, Deffayet:2002fn, Kiritsis:2002ca}. In this regard, Randall and Sundrum (RS) \cite{Randall:1999vf} proposed a model where a single $3$-brane is embedded in a $5$-dimensional ($5D$) anti de Sitter (AdS) bulk. AdS space-time is dual to a conformal field theory (CFT) living at its boundary through the so called AdS/CFT correspondence. The AdS/CFT correspondence \cite{Maldacena:1997re}, which is also a concrete illustration of the holographic principle, claims, in its original formulation, that classical $5D$ gravity in an AdS space-time is equivalent to a CFT on its boundary (For a review see \cite{Aharony:1999ti}).  More generally, the idea of holography proposes that the gravitational dynamics on higher dimensions may be understood from the gauge field theory on a lower dimension. In \cite{jcap} the holographic duality and the second RS model are combined to establish the cosmological brane-bulk energy exchange and the $4D$ description of the model.\par

Furthermore, constraints on the inflationary parameters in the context of the holographic duality have been studied for a universe filled by a scalar field in \cite{Lidsey:2005nt} and for a universe filled by a tachyon field in \cite{Bouabdallaoui:2016izz}. It was found that the holographic duality may describe the inflationary era and predicts the appropriate inflationary parameters comparing to the observational data. Our motivation has arisen from these investigations by modifying the standard brane-world scenario with the inclusion of an induced gravity (IG) effect added in the
brane action. The IG correction, which has been studied by many authors \cite{Kofinas:2001es, Deffayet:2000uy, Kiritsis:2002ca, Maeda:2003ar, Papantonopoulos:2004bm, BouhmadiLopez:2004ax, Bouhmadi-Lopez:2013gqa}, can be considered as a quantum correction coming from the bulk gravity and its coupling matter living on the brane \cite{Collins:2000yb, Dvali:2000hr}.\par

In this study, we will deal with holographic brane-world model modified by the IG correction. In particular, we will be interested in the primordial inflationary era by studying the inflationary parameters in the slow roll regime for two types of fields. Firstly, we will carry with a model driven by scalar field as responsible for cosmological inflation. Scalar fields which are known as the simplest form of matter are able to explain a wide range of complex phenomena for models of the early universe \cite{Guth:1980zm, Huey:2001ae} as well as for those of the late-time acceleration \cite{Caldwell:1999ew,Copeland:2006wr}. The idea underlying the scalar field inflationary scenario is that there exists a scalar field which is subject to the slow-roll approximation where the kinetic energy of the scalar field remains sufficiently small compared to its potential energy. Secondly, we will take into account the possibility that inflation may be driven by a tachyon field. As soon as tachyon fields are considered to play a significant role for early inflation phase, plenty of works have been done in brane-world cosmology \cite{BCT, BCT2}, D-branes inflation \cite{Sen:2002in}, multi tachyon fields \cite{Piao:2002vf}, warm inflationary model \cite{Kamali:2015yya, WIT} and k-inflation \cite{KI}.\par

In order to discriminate between the large number of inflationary models and to incorporate the inflationary scenario in the holographic setup, we compare the theoretical prediction of the spectral index of curvature perturbations with observations, but this still not sufficient to identify the best model of inflation. Therefore, further analysis are required such as the consistent behavior of the spectral index versus the tensor to scalar ratio or versus the running of the spectral index. This analysis can potentially provide further important informations to reduce the number of inflation models.\par

Current constraints from the Planck data \cite{Akrami:2018odb} suggest an upper limit of the tensor to scalar ratio $r < 0.1$ (Planck alone) at 95\% confidence level (C.L.), a value of the spectral index $n_{s}=0.9649\pm 0.0042$ and a value of the running of the spectral index $\alpha_{s}=-0.0045\pm0.0067$ quoted to $68$\% CL.\par

In the present paper, we take both holographic cosmology and IG curvature effects into consideration in order to study inflationary parameters of
the universe in which its dynamic is driven by a scalar field rolling down by a quadratic potential and a tachyon scalar field rolling down by an exponential potential. By comparing with the latest observational Planck data \cite{Akrami:2018odb}, we constraint the model's parameters.\par

The outline of this paper is the following. In Sec. II, we present the setup of an IG correction within the holographic cosmology. In this section, we introduce the background and the perturbative inflationary parameters  of both scalar and  tachyon fields respectively. In Sec. III, we confront our theoretical predictions with Planck2018 data. Finally, we present our conclusions in Sec. V.\par

\section{Setup}
We consider a generalized Randall Sundrum model with an IG term localized on the brane and its action is given by the following expression \cite{BouhmadiLopez:2004ax}
\begin{widetext}
 \begin{align}
\label{action}
S=\int_{bulk}d^{5}x\sqrt{-g^{\left( 5\right) }}\left( \frac{1}{2\kappa_{5}^{2}}R_{5}-\Lambda
_{5}\right)-\int_{brane}d^{4}x\sqrt{-\overset{}{g}}
\left(\frac{\gamma}{2 \kappa_{4}^{2}} R+\mathcal{L}_{x}-\Lambda_{4}\right),
\end{align}
 \end{widetext}
where $\kappa_{5}^{2}$ is the 5D gravitational constant, $R_{5}$ is the Ricci scalar of the five-dimensional metric $g^{(5)}$ and $\Lambda_{5}$ is the
bulk cosmological constant. In the brane action, $R$ is the Ricci scalar of the induced metric $g$, $\Lambda_{4}$ is the brane tension, $\gamma$ is a
dimensionless constant controlling the strength of the IG correction, with $\gamma=0$ giving the RS model. We require $0\leq\gamma<1$ in order to have a positive effective gravitational coupling constant for the low and high energy limits \cite{BouhmadiLopez:2004ax}. $\mathcal{L}_{x}$ is the Lagrangian density for the scalar field, i.e. $x=\phi$, or the tachyon field, i.e. $x=T$.\par

As a background model universe we consider a spatially flat isotropic and homogeneous Friedmann-Robertson-Walker (FRW) metric with vanishing spatial curvature and cosmological constant. In the holographic setup with an IG correction the Friedmann equation becomes \cite{Bargach:2018ujh}\footnote{Note that this equation can be linked to the one in Ref. \cite{Bargach:2018ujh} by setting $f=\frac{\gamma}{2\kappa_{4}^{2}}$.}
\begin{equation}\label{friedmanmod}
 H^{2} = \frac{1+\gamma}{4 c \kappa_{4}^{2}} \left [ 1\pm\sqrt{1- \frac{\rho_{x}}{\rho_{max}}}\right],
\end{equation}
where $c$ is the conformal anomaly coefficien, $\rho_{x}$ is the energy density, $\rho_{max} = \frac{3 (\gamma+1)^{2}}{8c \kappa_{4}^{4}}$ and the sign ($\pm$) shows the existence of two branches of solution. We recover the standard form of the Friedmann equation at low-energy limit, $\rho_{x}\ll \rho_{max}$, for the limit $\gamma\rightarrow 0$ and for the negative branch. Hereafter only this branch will be considered.\par
\subsection{Scalar field inflation}

\subsubsection{Background parameters}
In this subsection, we consider that inflation is driven by a scalar field, $\phi$. The Lagrangian density of the scalar field localized on the brane, is defined as
\begin{equation}\label{lphi}
  \mathcal{L_{\phi}} =\frac{1}{2} g^{\mu \nu }\nabla _{\mu }\phi \nabla _{\upsilon }\phi -V(\phi),
\end{equation}
and the Friedmann equation \eqref{friedmanmod} yields
\begin{equation}\label{friedmannscalar}
 H^{2} = \frac{1+\gamma}{4 c \kappa_{4}^{2}} \left [ 1-\sqrt{1- \frac{\rho_{\phi}}{\rho_{max}}}\right],
\end{equation}
where the energy density has the form
\begin{equation}\label{rhoscalar}
  \rho_{\phi}= \frac{1}{2}\dot{\phi}^{2} + V,
\end{equation}
and the equation of motion takes the following form
\begin{equation}\label{motionfield}
 \ddot{\phi}+3H\dot{\phi}+V_{,\phi}=0,
 \end{equation}
where a dot corresponds to a derivative with respect to the cosmic time and we use the subscript ($,\phi$) to denote a derivative with respect
to $\phi$.\par

During the inflationary epoch and assuming a slow-roll expansion, i.e. $\dot{\phi}^{2}<<V$ and $\ddot{\phi}<<3H\dot{\phi}$, the Friedmann equation
\eqref{friedmannscalar} can be rewritten using \eqref{rhoscalar} as
\begin{equation}\label{friedmanslow}
  H^{2}=\frac{1+\gamma}{4 c \kappa^{2}_{4}}\left[1-\sqrt{1-U}\right],
\end{equation}
where $U\equiv V/V_{max}$ is a dimensionless parameter and $V_{max}=\frac{3 (\gamma+1)^{2}}{8c \kappa_{4}^{4}}$, the standard cosmology is recovered for $U\ll 1$ and $\gamma=0$. Also, the equation of motion \eqref{motionfield} reduces to
\begin{equation}\label{motionslow}
  \dot{\phi}\simeq -\frac{V_{,\phi}}{3H}.
\end{equation}

The slow-roll parameters defined as
\begin{equation}\label{epsilon}
  \epsilon\equiv-\frac{\dot{H}}{H^{2}}, \qquad \eta\equiv\frac{V_{,\phi\phi}}{3H^{2}},
\end{equation}
can be rewritten by using equations \eqref{friedmanslow} and \eqref{motionslow} as
\begin{equation}\label{eps}
  \epsilon\simeq\frac{1}{2\kappa_{4}^{2}}\left(\frac{V_{,\phi}}{V}\right)^{2}C_{\gamma,c}^{(a)},
\end{equation}
and
 \begin{equation}\label{eta}
   \eta\simeq \frac{1}{\kappa_{4}^{2}}\left(\frac{V_{,\phi\phi}}{V}\right)C_{\gamma,c}^{(b)},
\end{equation}
where $C_{\gamma,c}^{(a)}$ and $C^{(b)}_{\gamma,c}$ denote correction terms to the standard  four dimension ($4D$) expressions. Their forms are given respectively by
 \begin{equation}\label{Ca}
  C^{(a)}_{\gamma,c}=\frac{(1+\gamma)(1+\sqrt{1-U})^{2}}{4\sqrt{1-U}},
 \end{equation}
 and
  \begin{equation}\label{cb}
  C^{(b)}_{\gamma,c}= \frac{(1+\gamma) (1+\sqrt{1-U})}{2}.
 \end{equation}
These correction terms depend on both effects, holographic cosmology and IG coupling. We can notice that at the low energy limit ($V\ll V_{max}$) and for $\gamma\rightarrow 0$ the correction terms reduce to one and the standard slow roll parameters are recovered.\\

  The number of e-folds during inflation is given by
 \begin{equation}\label{efolds}
 N=  \int_{t_{i}}^{t_{f}}H dt,
 \end{equation}
 which in the slow-roll approximation can be written as
 \begin{equation}\label{efoldsslow}
  N\simeq -\frac{2 \kappa_{4}^{2}}{1+\gamma}\int_{\phi_{i}}^{\phi_{f}}\frac{1-\sqrt{1-U}}{U_{,\phi}}d\phi,
 \end{equation}
where $\phi_{i}$ denotes the value of the scalar field when the radius of the universe crosses the Hubble horizon during inflation and $\phi_{f}$ its value when the universe exits the inflationary phase.\par

\subsubsection{Perturbative parameters}
In this subsection, we explore the linear perturbation theory in the inflationary era driven by a scalar field. In the longitudinal gauge, the scalar metric perturbations of the FRW background are given by \citep{Bardeen:1980kt, Mukhanov:1990me}
\begin{equation}\label{gauge}
  ds^{2}=-(1+2\Phi)dt^{2}+a^{2}(t)(1-2\Psi)\delta_{ij}dx^{i}dx^{j},
\end{equation}
where $a(t)$ is the scale factor, $\Phi(t, x)$ and 	$\Psi(t, x)$ are the scalar perturbations.
The curvature perturbation on uniform density hypersurfaces, in terms of scalar field fluctuations on spatially flat hypersurfaces, is given by
$\zeta=H \delta\phi/\dot{\phi}$ where the field fluctuations at Hubble crossing ($k=aH$) and within
the slow-roll limit are given by $<\delta\phi^{2}>=(H/2\pi)^{2}$, as the equation of motion is unaffected by the brane-world model under study. Consequently, the power spectrum of the curvature perturbations is given by
 \cite{Lidsey:2005nt}
\begin{equation}\label{as}
  A_{s}^{2}=\frac{4}{25}<\zeta^{2}> = \frac{1}{25\pi^{2}}\frac{H^{4}}{\dot{\phi}^{2}}.
\end{equation}
In our model and within the slow-roll approximation, we find
\begin{equation}\label{asmodi}
  A_{s}^{2}\simeq \frac{\kappa_{4}^{6}}{75
   \pi^{2} } \frac{ V^{3}}{V_{,\phi}^{2}} \frac{1}{(C_{\gamma,c}^{(b)})^{3}}.
\end{equation}
The scalar spectral index to first order in the slow-roll parameters is described by the spectral tilt
\begin{equation}\label{ns}
  n_{s}-1=\frac{dlnA_{s}^{2}}{dlnk} \simeq -6 \epsilon +2 \eta.
\end{equation}

Furthermore, the amplitude of gravitational waves are bound to the brane at long-wavelengths and they are decoupled, to a first order, from the matter perturbations. Hence, at large scales, the amplitude is obtained by the Hubble rate when each mode exits the Hubble scale during inflation\footnote{Here, the amplitude of gravitational waves is assumed to be the same as the 4D result, i.e. we neglect the correction to standard 4D general relativity ($F_{\gamma}\simeq 1$), see \cite{BouhmadiLopez:2004ax} for the full expression of $F_{\gamma}$.} \cite{Maartens:1999hf}. It is then sufficient to use the tensor perturbations amplitude of a given mode, at the Hubble crossing, given by\footnote{This formula is also valid for a tachyon field \cite{Nozari:2013mba}.} \cite{ Nozari:2012cy}
\begin{equation}\label{At}
  A_{T}^{2}= \frac{4\kappa_{4}^{2}}{25 \pi} H^{2} \mid_{k=aH}.
\end{equation}
In our model and within the slow-roll approximation, we find
\begin{equation}\label{Atmod}
   A_{T}^{2}\simeq \frac{ 4\kappa_{4}^{4}}{75 \pi}  \frac{V}{C_{\gamma,c}^{(b)}}.
\end{equation}
The tensor spectral index is given by
\begin{equation}\label{nT}
  n_{T}= \frac{dln A_{T}^{2}}{dlnk},
\end{equation}
which can be expressed in terms of the slow-roll parameters as
\begin{equation}\label{nTepsilon}
  n_{T}\simeq -2 \epsilon.
\end{equation}

Another important and useful inflationary parameter which can be compared with the observation is the tensor-to-scalar ratio
\begin{equation}\label{r}
  r\equiv \frac{A_{T}^{2}}{A_{S}^{2}} \simeq   8\pi\epsilon\; (1+\gamma) \sqrt{1-U}.
\end{equation}

\subsection{Tachyon field}

\subsubsection{Background parameters}
We now consider the model with a tachyon field. The Lagrangian density for a tachyon field can be written as
\begin{equation}\label{lagrangiantachyon}
  \mathcal{L_{T}} =-\sqrt{1-\nabla _{\mu }T \nabla _{\mu }T}\; V(T),
\end{equation}
where $T$ is the tachyon field and $V(T)$ its potential. The Friedmann equation \eqref{friedmanmod} for a model with a tachyon field is obtained as follows
\begin{equation}\label{fr}
   H^{2} = \frac{1+\gamma}{4 c \kappa_{4}^{2}} \left [ 1- \sqrt{1- \frac{\rho_{T}}{\rho_{max}}}\right],
\end{equation}
where $\rho_{T}$ is the energy density for the tachyon field which can be expressed as \cite{Sen:2002in}
\begin{equation}\label{ene}
  \rho_{T}=\frac{V(T)}{\sqrt{1-\dot{T}^{2}}}.
  \end{equation}

The equation of motion of a tachyon field propagating on the brane is \cite{ArmendarizPicon:1999rj}
\begin{equation}\label{motion}
  \frac{\ddot{T}}{1-\dot{T}^{2}}+3H\dot{T}+\frac{V_{,T}}{V(T)}=0,
\end{equation}
where a dot corresponds to a derivative with respect to the cosmic time and $V_{,T}=dV/dT$.\par

During the inflationary era and in the slow-roll approximation, i.e. $\dot{T}^{2}<<1 $ and $ \ddot{T}<< 3H \dot{T} $; the Friedmann equation for the tachyon field takes the form of Eq. \eqref{friedmanslow} and the equation of motion for tachyon field Eq. \eqref{motion} reduces to
\begin{equation}\label{slo}
  3H\dot{T}\simeq -\frac{V_{,T}}{V(T)}.
\end{equation}

The slow roll parameters of the tachyon field, denoted by a subscript $T$, take the following form
\begin{equation}\label{epsilon}
  \epsilon_{T}\simeq \frac{1}{2\kappa_{4}^{2}}\frac{V_{,T}^{2}}{V(T)^{3}} C^{(a)}_{\gamma,c},
\end{equation}
and
\begin{equation}\label{et}
  \eta_{T}\simeq \frac{1}{\kappa_{4}^{2}}\frac{V_{,TT}}{V(T)^{2}}C^{(b)}_{\gamma,c},
\end{equation}
where the correction terms $C^{(a)}_{\gamma,c}$ and $C^{(b)}_{\gamma,c}$ are the same as those of the scalar field and are given respectively by Eqs.\eqref{Ca} and \eqref{cb}.\par

\subsubsection{Perturbative parameters}
In this subsection, we study the cosmological perturbations in the slow roll regime for the tachyon field. The power spectrum of the curvature perturbations is given by\footnote{Note that this standard 4D formula is expected to remain true
when induced gravity corrections are included, as the equation of motion for a tachyonic field is unaffected by the brane-world model.} \cite{Nozari:2013mba}
\begin{equation}\label{as}
  A_{S}^{2}=\frac{1}{2\pi^{2}}\frac{H^{4}}{\dot{\phi}^{2}}\frac{1}{V}.
\end{equation}
Using the slow roll condition, Eq. \eqref{as} takes the following form
\begin{equation}\label{a}
  A_{S}^{2}\simeq\frac{\kappa_{4}^{6}}{12\pi^{2}}\frac{V^{4}}{V_{,T}^{2}} (C_{\gamma,c}^{(b)})^{-3}.
\end{equation}
The scalar spectral index within the slow-roll approximations is given by
\begin{equation}\label{nsT}
  n_{s}-1\simeq-6 \epsilon_{T} +2\eta_{T}.
\end{equation}

The amplitude of tensor perturbations is given by \cite{Nozari:2013mba}
\begin{equation}\label{tensor}
  A_{T}^{2}=\frac{4 \kappa_{4}^{2}}{25 \pi}H^{2},
\end{equation}
 and within the slow roll limit it takes the following form
 \begin{equation}\label{te}
    A_{T}^{2}\simeq\frac{4  \kappa_{4}^{4}}{75 \pi}\frac{V}{C^{(b)}_{\gamma,c}}.
 \end{equation}
The spectral index related to the tensor perturbation is defined by $n_{T}=dlnA_{T}^{2}/dlnk$, and from Eq. \eqref{tensor} we find
\begin{equation}\label{ntT}
  n_{T} \simeq -2 \epsilon_{T}.
\end{equation}
 The tensor to scalar ratio of the tachyonic field is as follows
 \begin{equation}\label{r}
   r\simeq\frac{32\pi}{25} \epsilon_{T}(1+\gamma) \sqrt{1-U}.
 \end{equation}

\section{Observational constraints}

\subsection{Constraints on model's parameters}
In this subsection, we are interested in constraining the holographic cosmology using observational data \cite{Akrami:2018odb}. From the above equations we note that the slow roll parameters, the spectral index and the tensor to scalar ratio are modified by a correction terms for both scalar and tachyon fields. These correction terms can be evaluated as a function of the dimensionless parameter $U$ for a given set of parameters $c$ and $\gamma$. From Eqs. \eqref{friedmanmod}, \eqref{At} and using the definition $r=A_{T}^{2}/A_{S}^{2}$ a relationship between our model's parameters is obtained as
\begin{equation}\label{U}
  U= \frac{25 \pi A_{s}^{2} c}{(1+\gamma)} r \left(2-\frac{25 \pi A_{s}^{2} c}{(1+\gamma)} r\right).
\end{equation}
The same relationship have been obtained before for $\gamma=0$, i.e. without the effect of an induced gravity, with a universe filled by a tachyon field in the context of holographic cosmology \cite{Bouabdallaoui:2016izz}.
\begin{figure*}
  \centering
\begin{tabular}{ccc}
     \subfloat[ $r=0.06$  \label{fig1a}]{\includegraphics[scale=0.5]{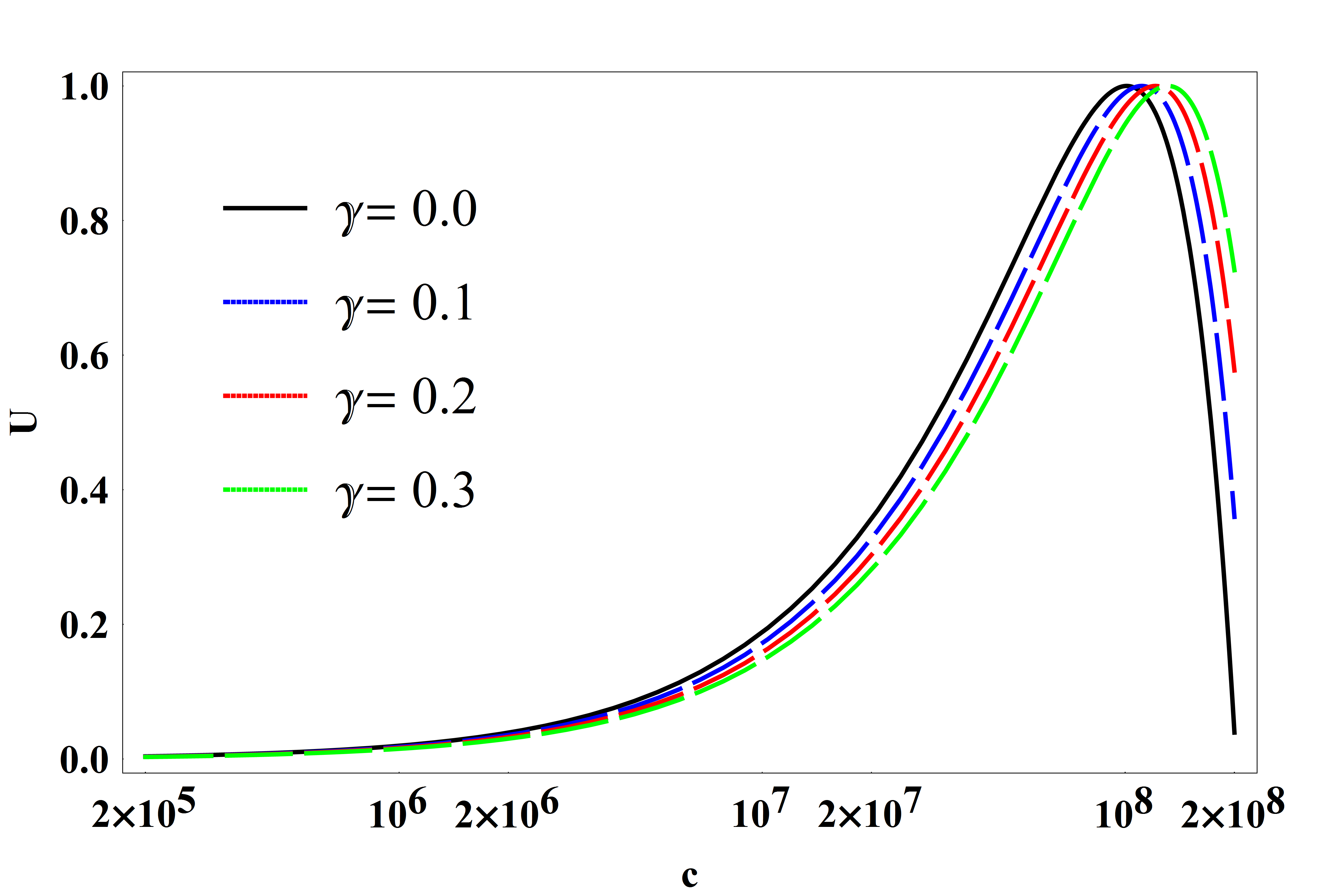}} &
     \subfloat[$\gamma=0.2$ \label{fig1b}]{\includegraphics[scale=0.5]{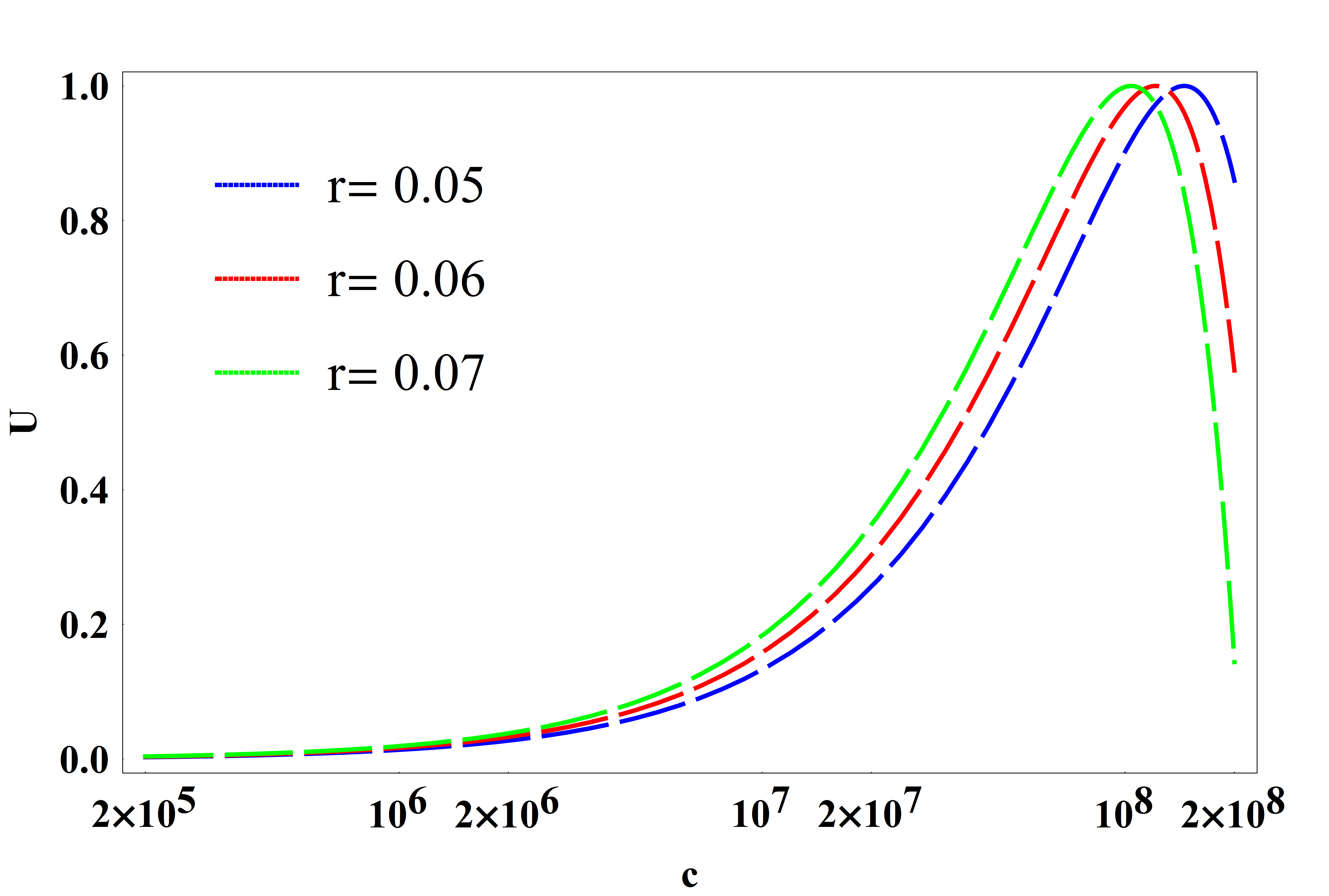}}&
     \subfloat[$r=0.06$  \label{fig1c}]{\includegraphics[scale=0.5]{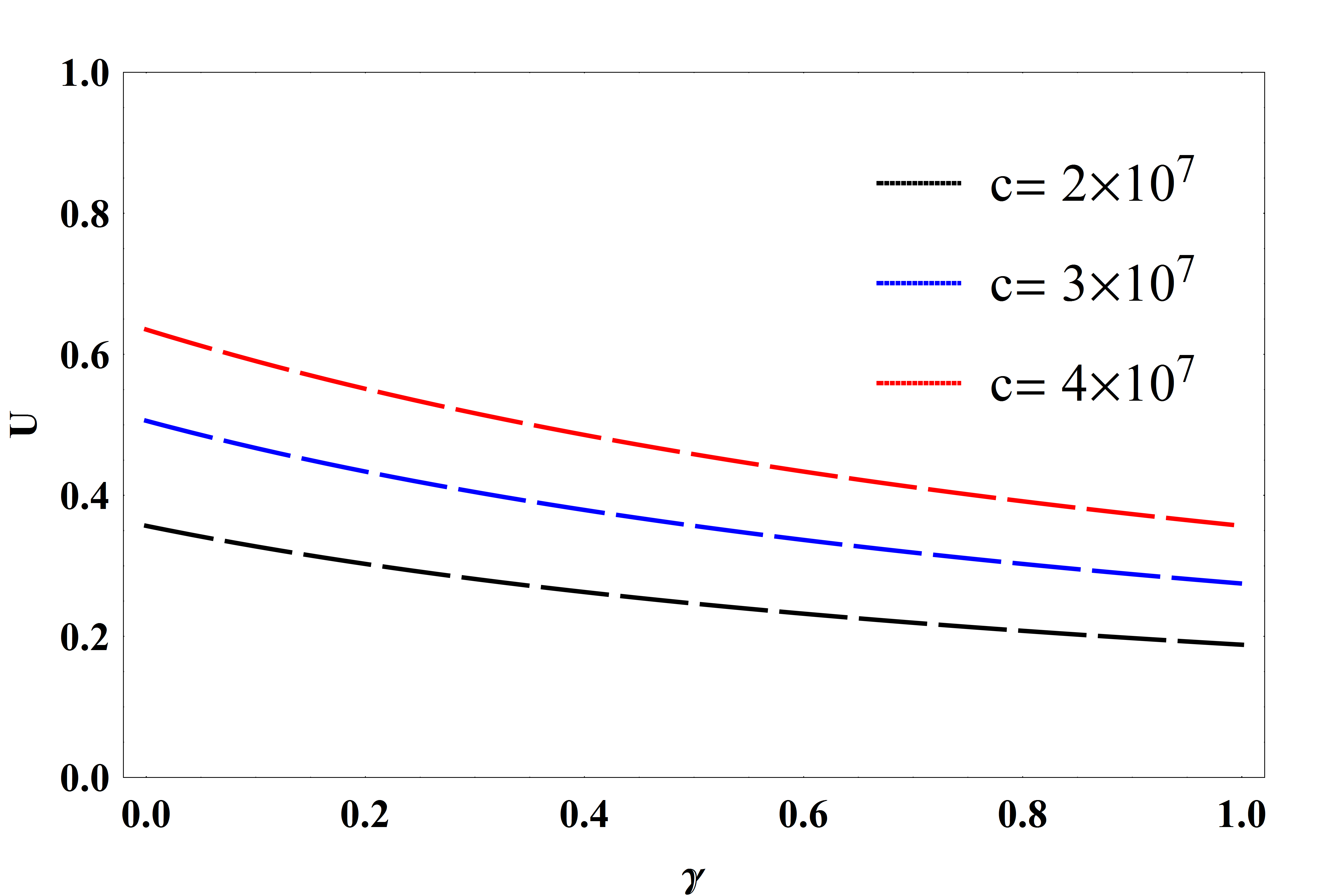}}\\
     \subfloat[$c=2\times10^{7}$ \label{fig1d}]{\includegraphics[scale=0.5]{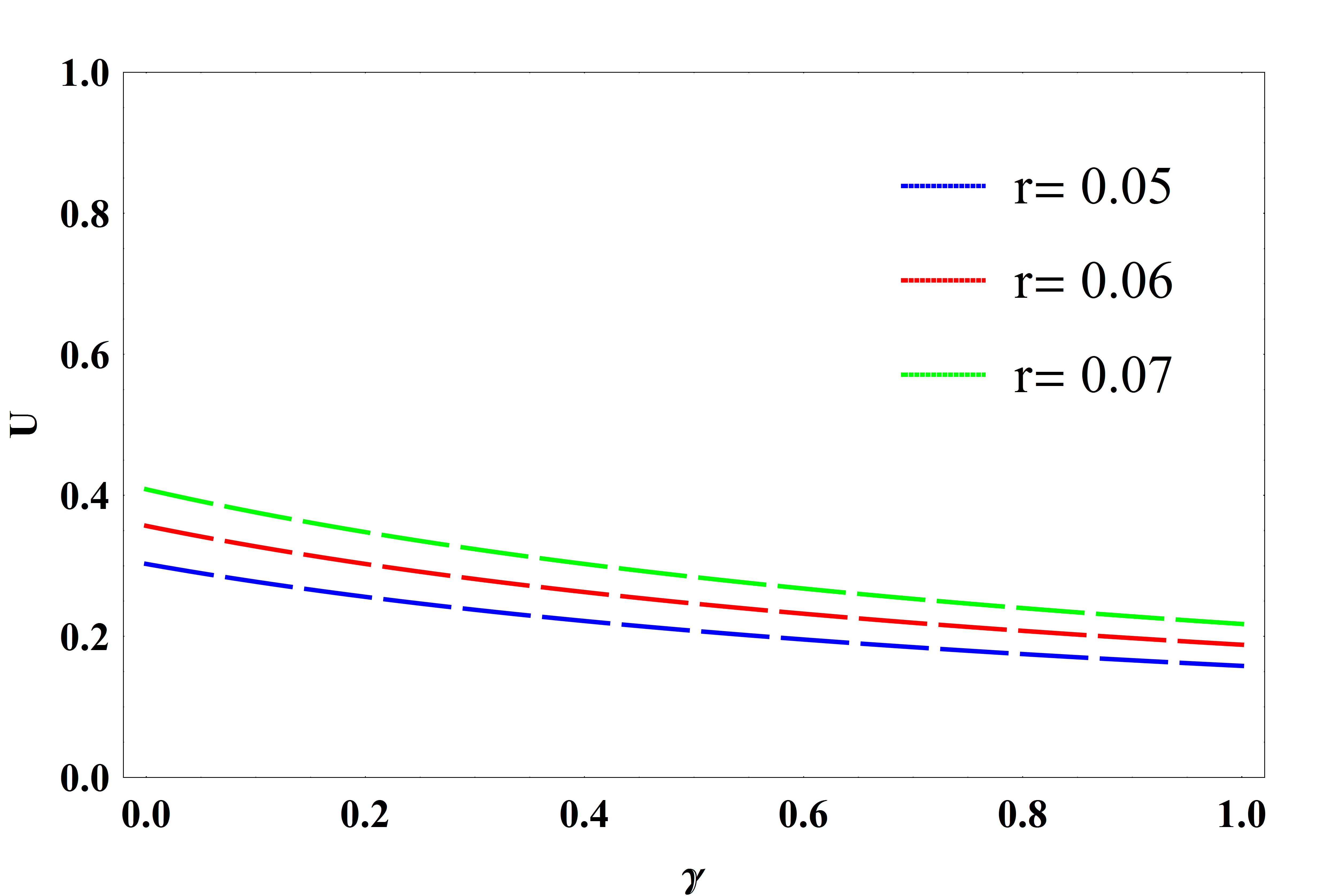}}&
     \subfloat[$\gamma=0.1$ \label{fig1e}]{\includegraphics[scale=0.5]{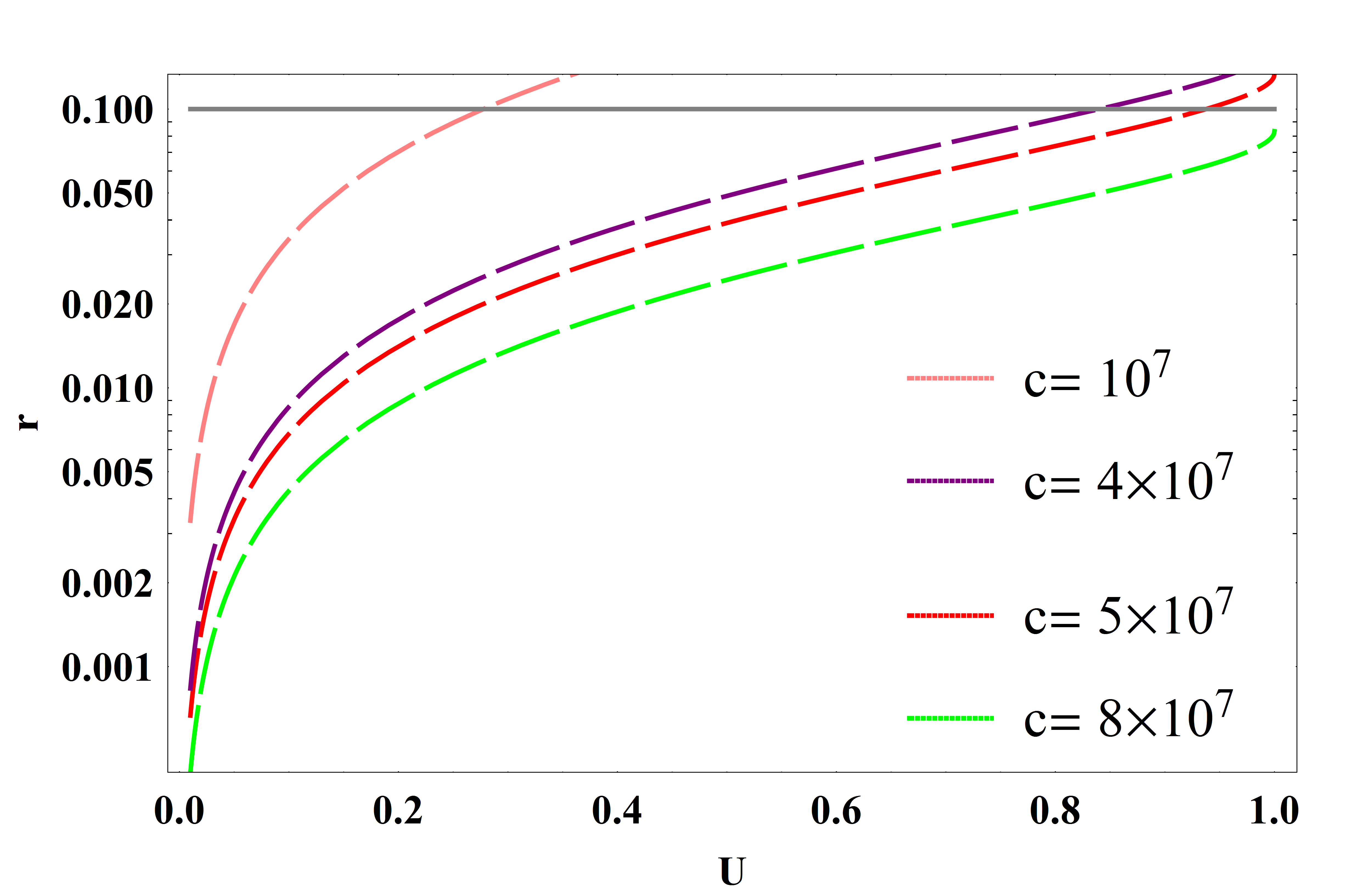}}&
     \subfloat[$c=5\times10^{7}$ \label{fig1f}]{\includegraphics[scale=0.5]{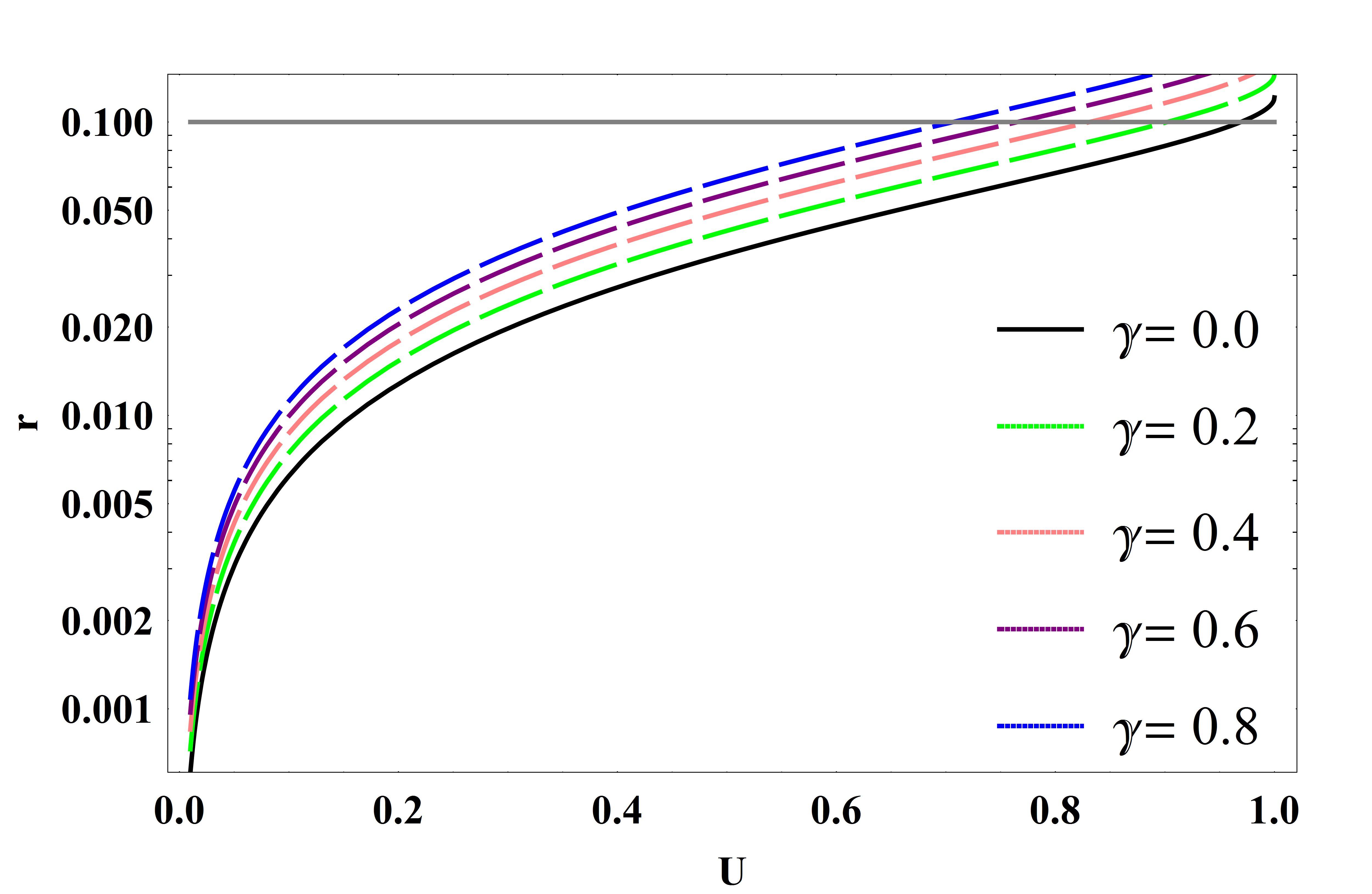}}\\
\end{tabular}
   \caption[]{The evolution of the dimensionless parameter $U$ versus the conformal anomaly coefficient $c$ and the IG parameter $\gamma$ respectively, and the tensor to scalar ratio $r$ versus the dimensionless parameter $U$. Here we have imposed the constraint on the power spectrum of scalar perturbations from Planck data \cite{Akrami:2018odb}: $A_{s}^{2}=2.101 \times 10^{-9}$. The horizontal gray line in (e) and (f) indicates the upper bound for the tensor-to-scalar ratio $r$ predicted by Planck.}\label{figure1}
\end{figure*}

\begin{figure*}
  \centering
  \begin{tabular}{ccc}
     \subfloat[ $r=0.06$ \label{fig2a}]{\includegraphics[scale=0.5]{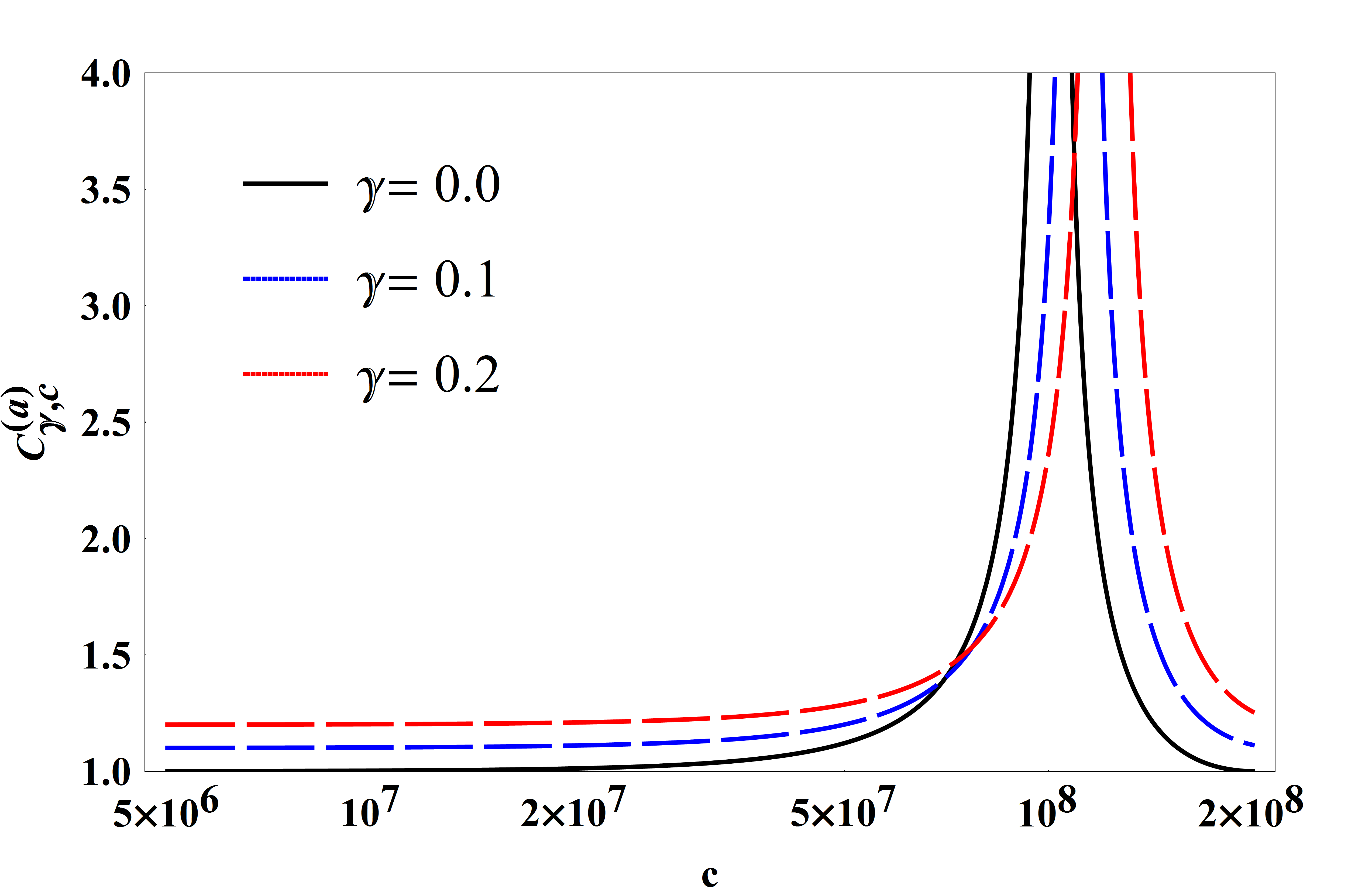}} &
     \subfloat[ $\gamma=0.1$ \label{fig2b}]{\includegraphics[scale=0.5]{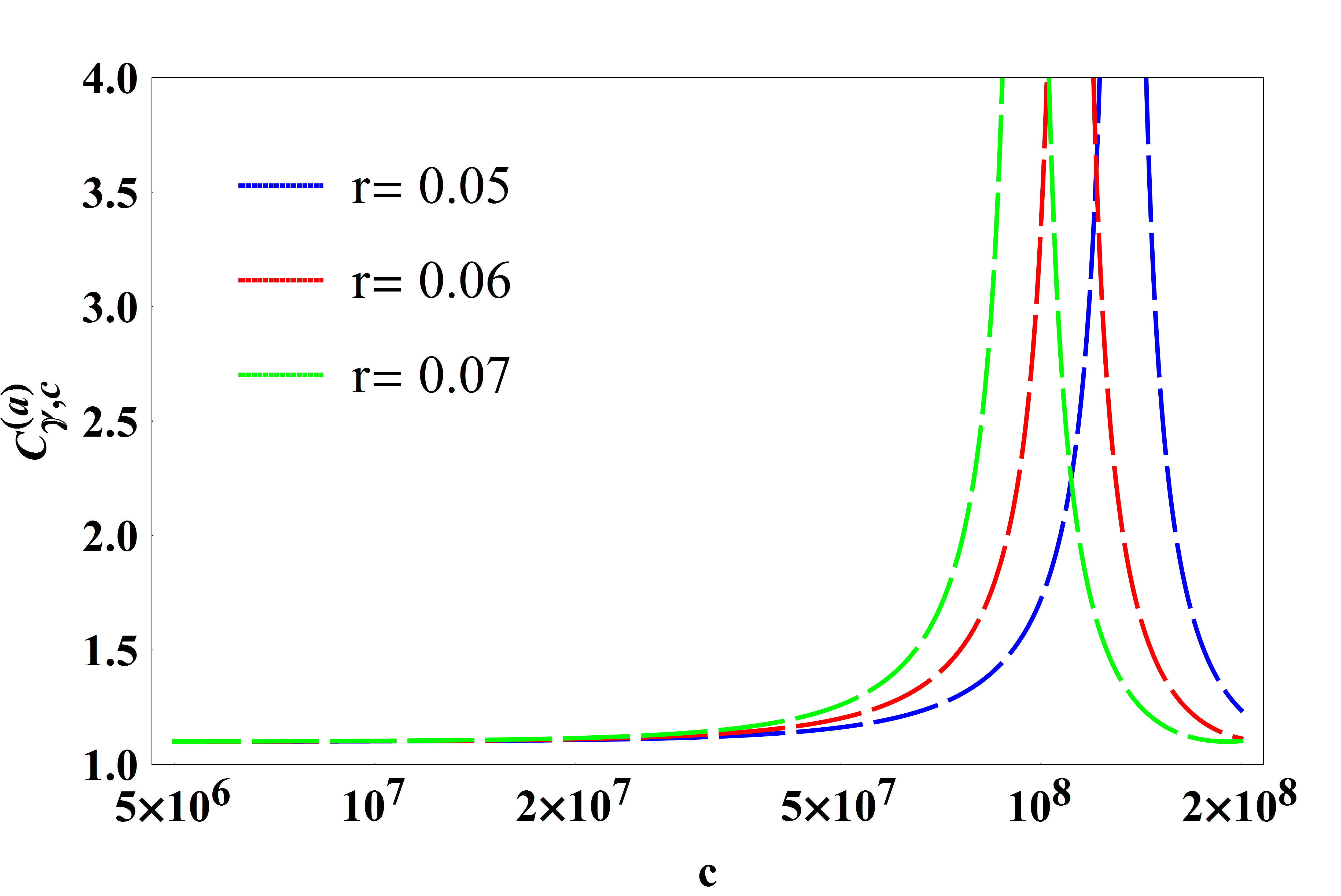}}&
     \subfloat[$r=0.06$ \label{fig2c}]{\includegraphics[scale=0.5]{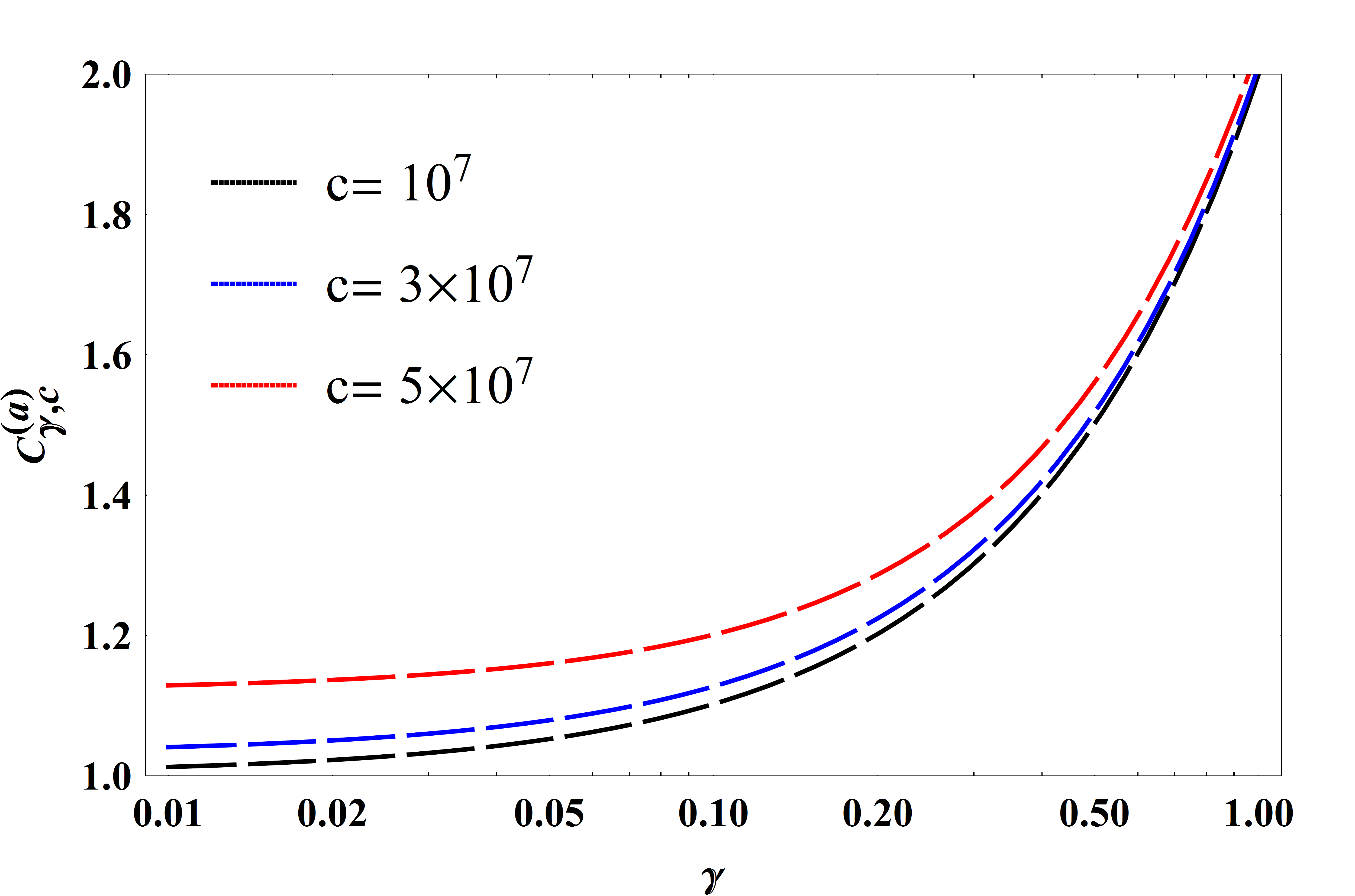}}\\
     \subfloat[$c= 5\times 10^{7}$ \label{fig2d}]{\includegraphics[scale=0.5]{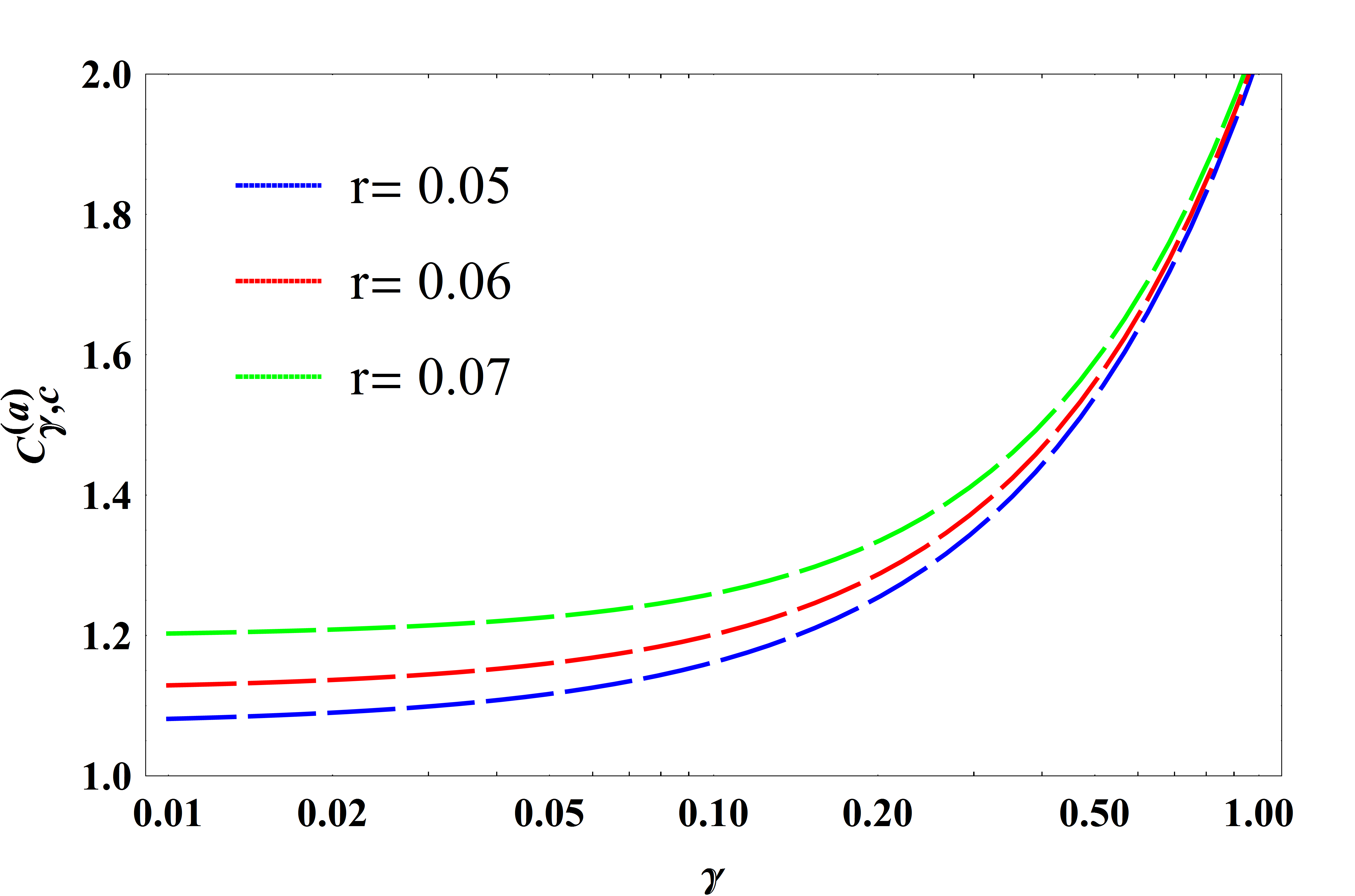}}&
     \subfloat[ $\gamma=0.2$ \label{fig2e}]{\includegraphics[scale=0.5]{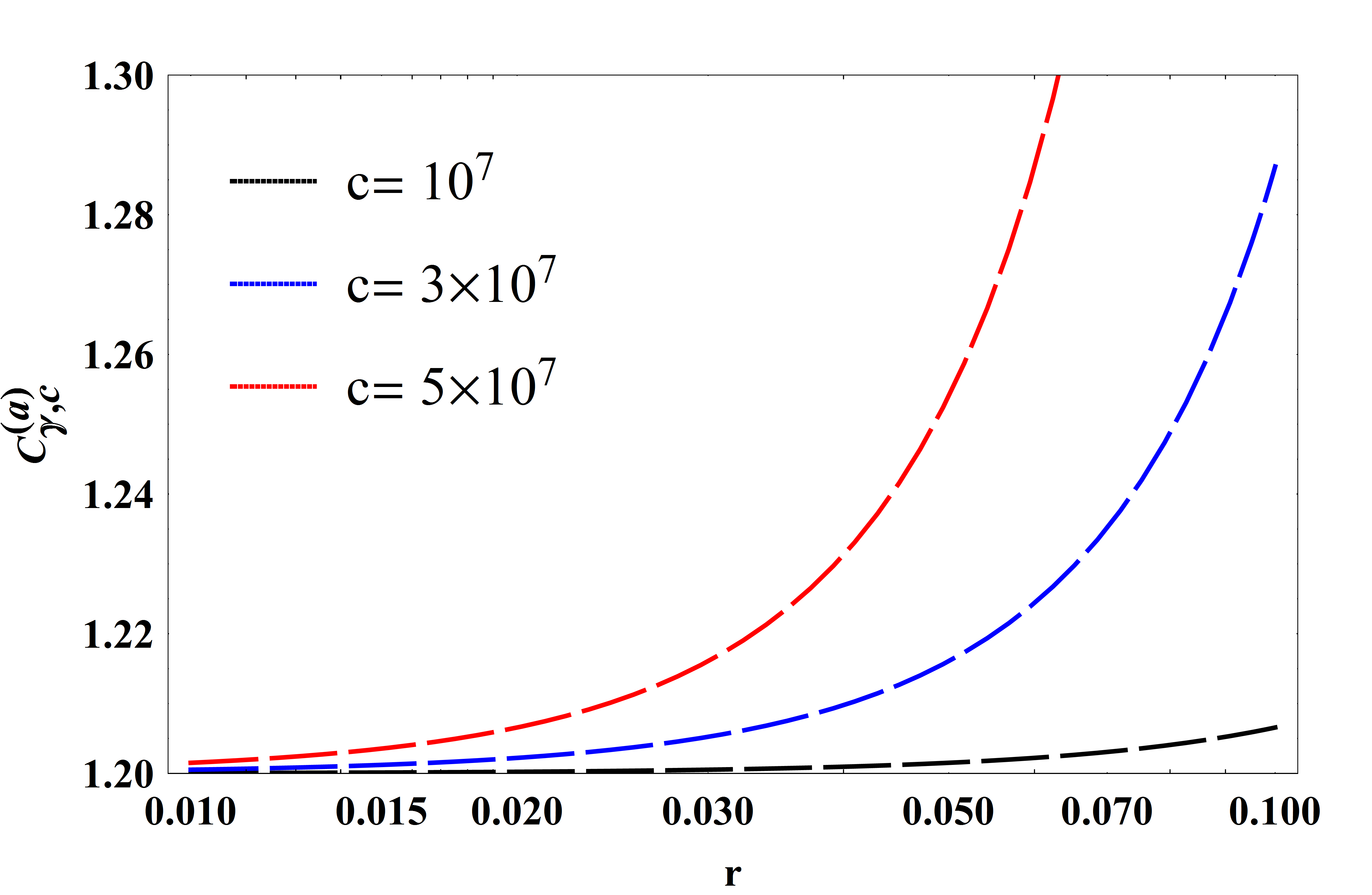}}&
     \subfloat[ $c= 6\times 10^{7}$ \label{fig2f}]{\includegraphics[scale=0.5]{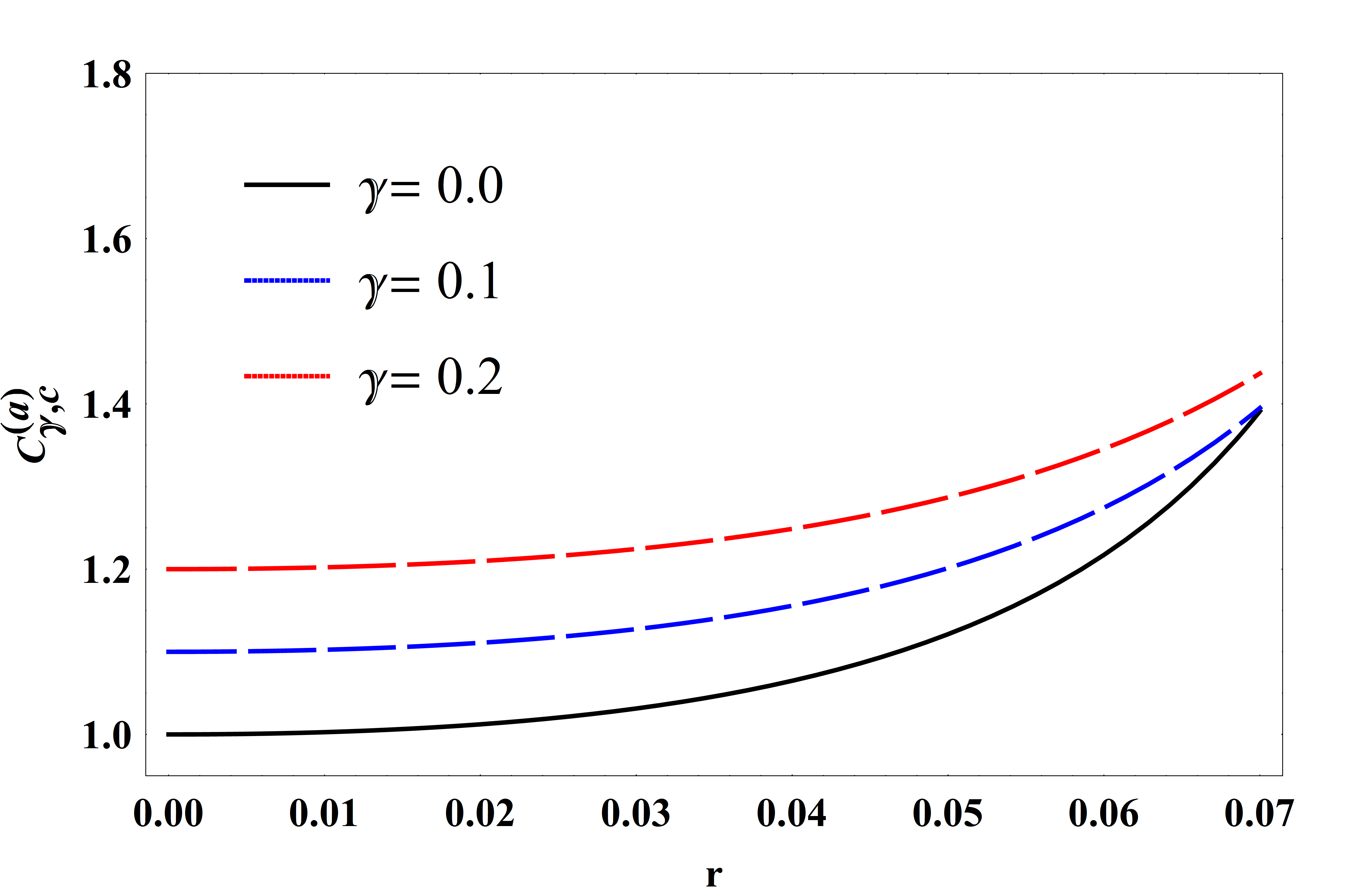}}\\

\end{tabular}
   \caption[]{Evolution of the correction term $C_{\gamma,c}^{(a)}$ versus the conformal anomaly coefficient $c$, the IG strength $\gamma$ and the tensor-to-scalar ratio $r$, respectively. Here we have imposed the constraint on the power spectrum of scalar perturbations from Planck data \cite{Akrami:2018odb}: $A_{s}^{2}=2.101 \times 10^{-9}$.}\label{figure2}
\end{figure*}
\begin{figure*}
  \centering
  \begin{tabular}{ccc}
     \subfloat[$r=0.06$ \label{fig3a}]{\includegraphics[scale=0.5]{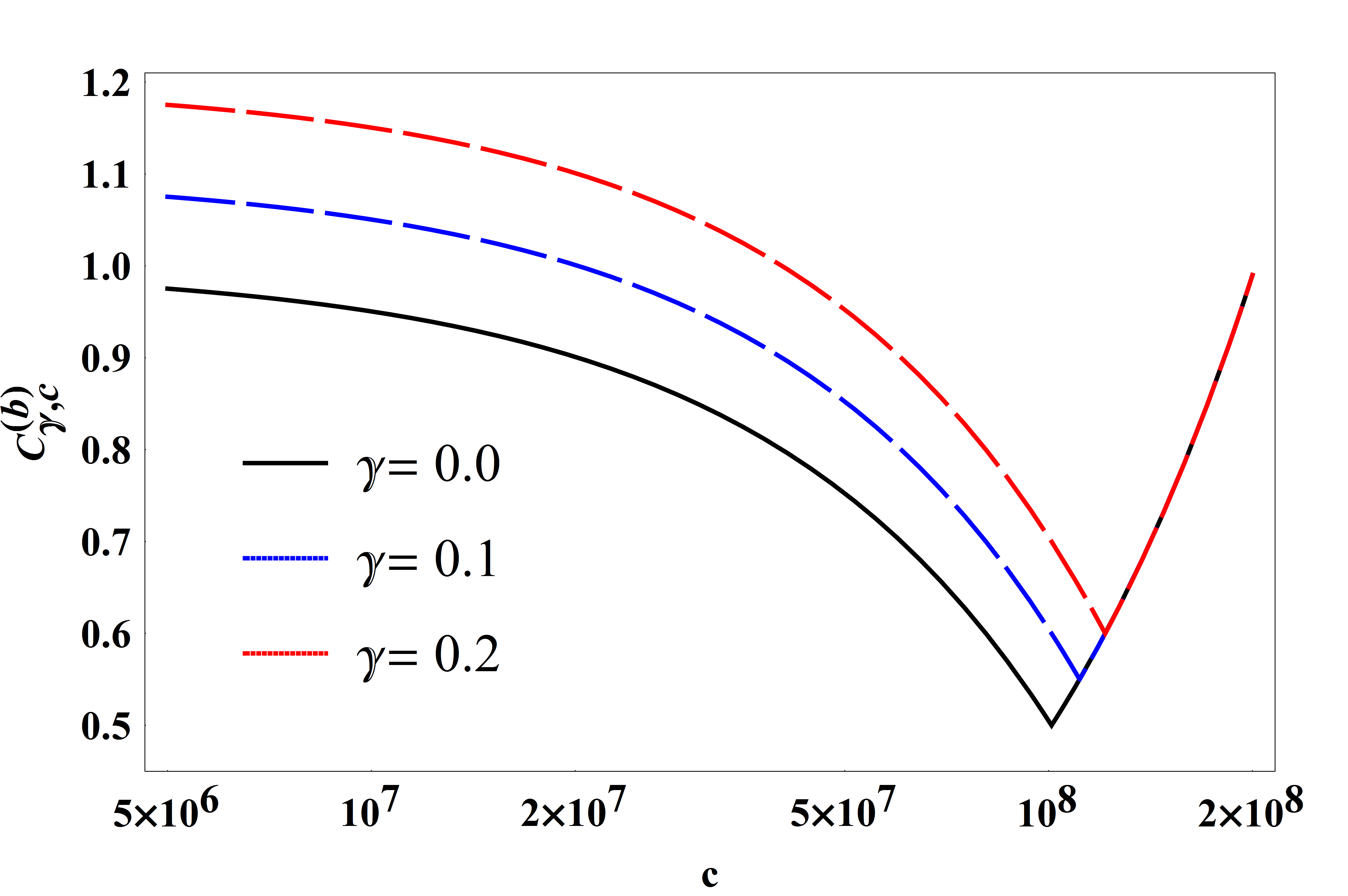}} &
     \subfloat[ $\gamma=0.1$\label{fig3b}]{\includegraphics[scale=0.5]{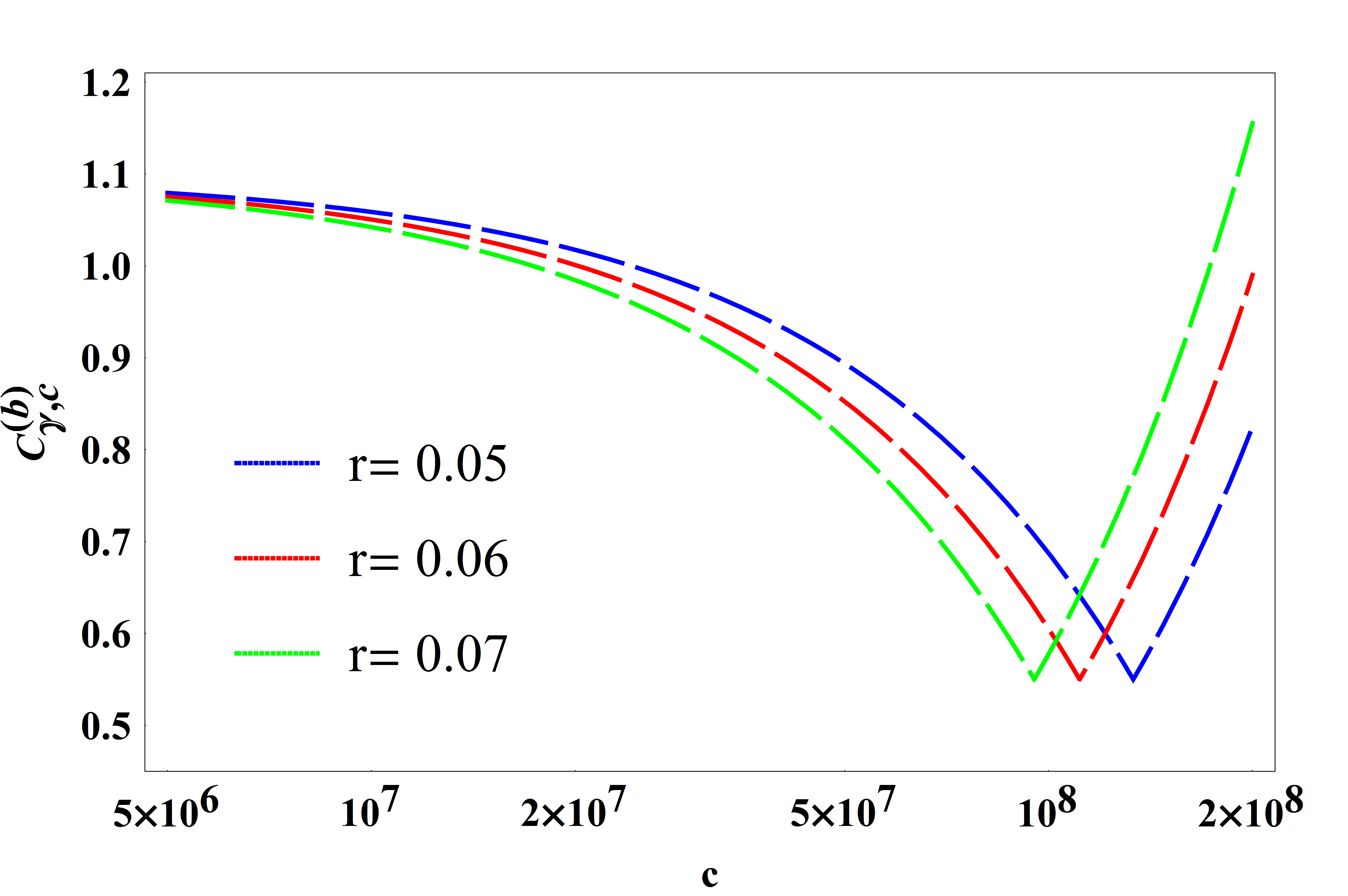}}&
     \subfloat[ $r=0.06$\label{fig3c}]{\includegraphics[scale=0.5]{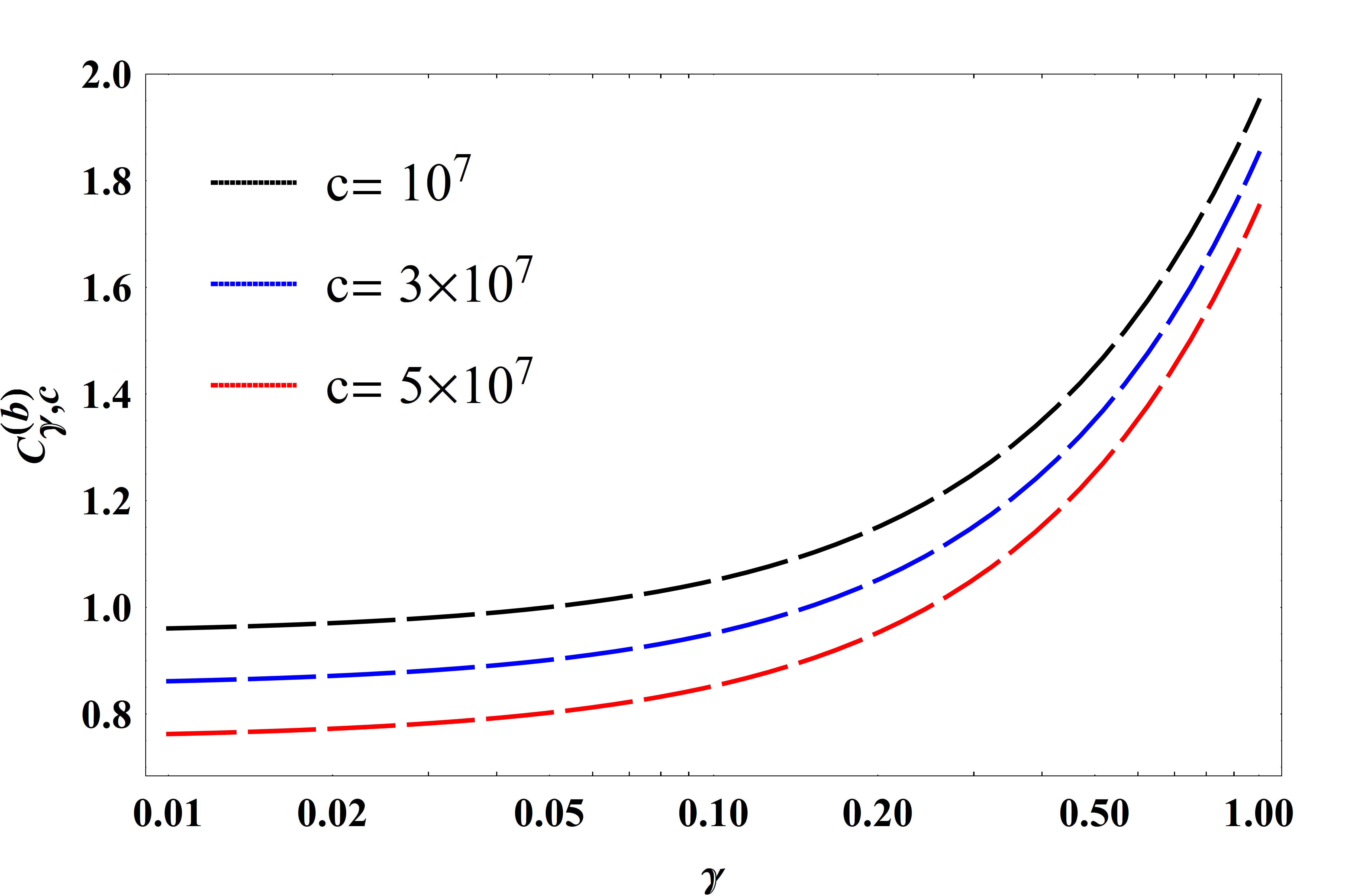}}\\
     \subfloat[$c= 5\times 10^{7}$ \label{fig3d}]{\includegraphics[scale=0.5]{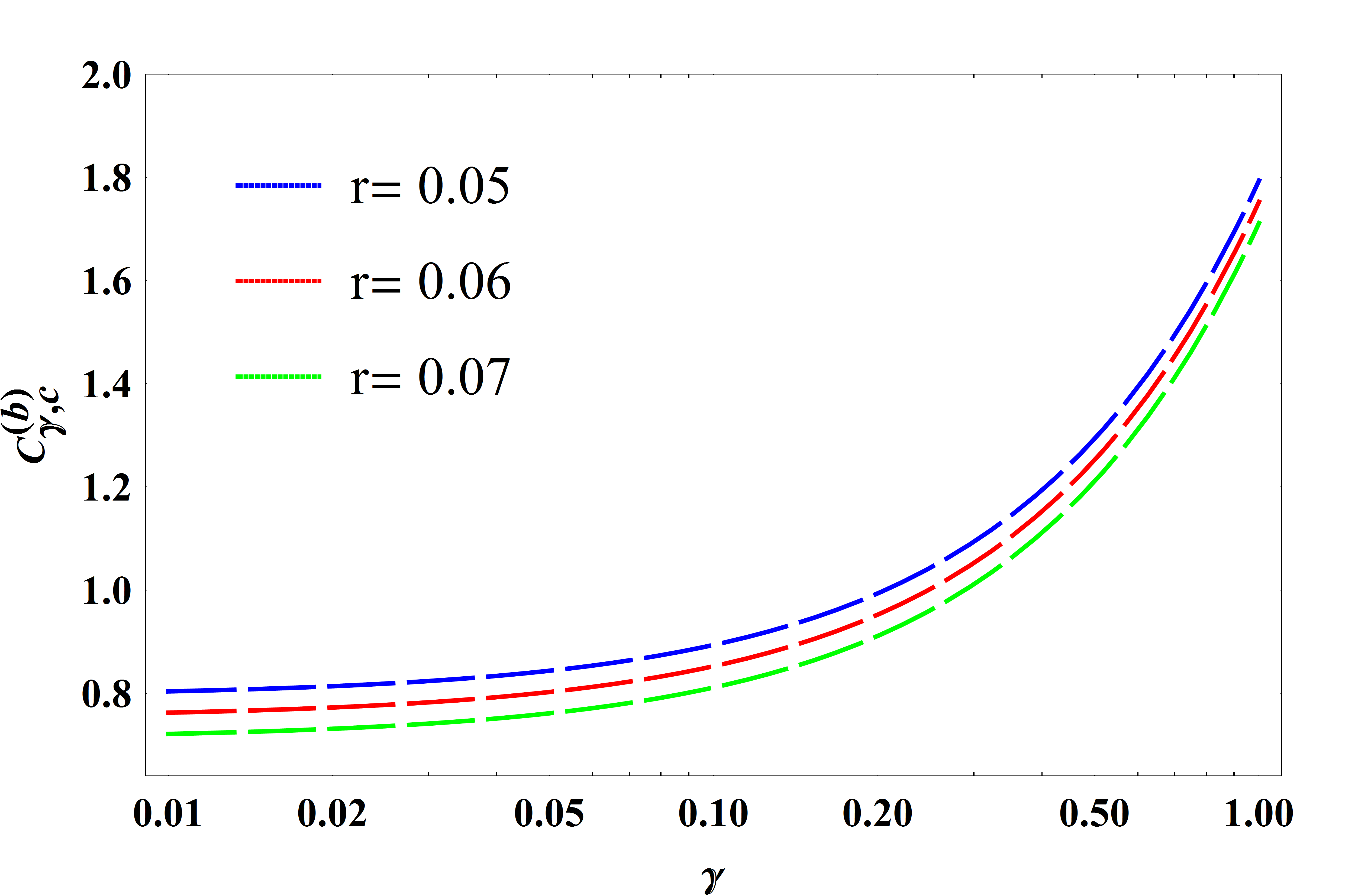}}&
     \subfloat[ $\gamma=0.2$\label{fig3e}]{\includegraphics[scale=0.5]{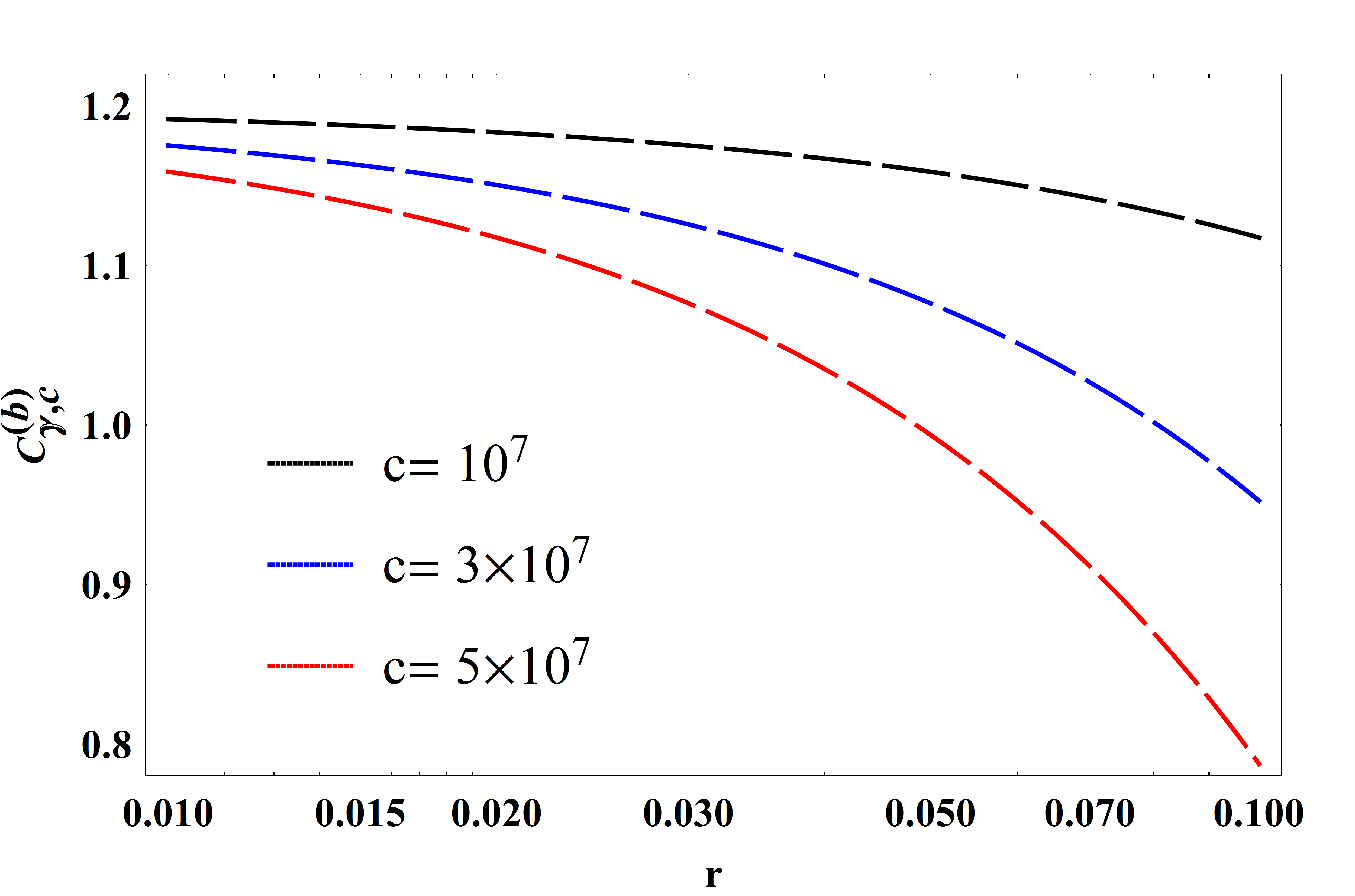}}&
     \subfloat[$c= 4\times 10^{7}$ \label{fig3f}]{\includegraphics[scale=0.5]{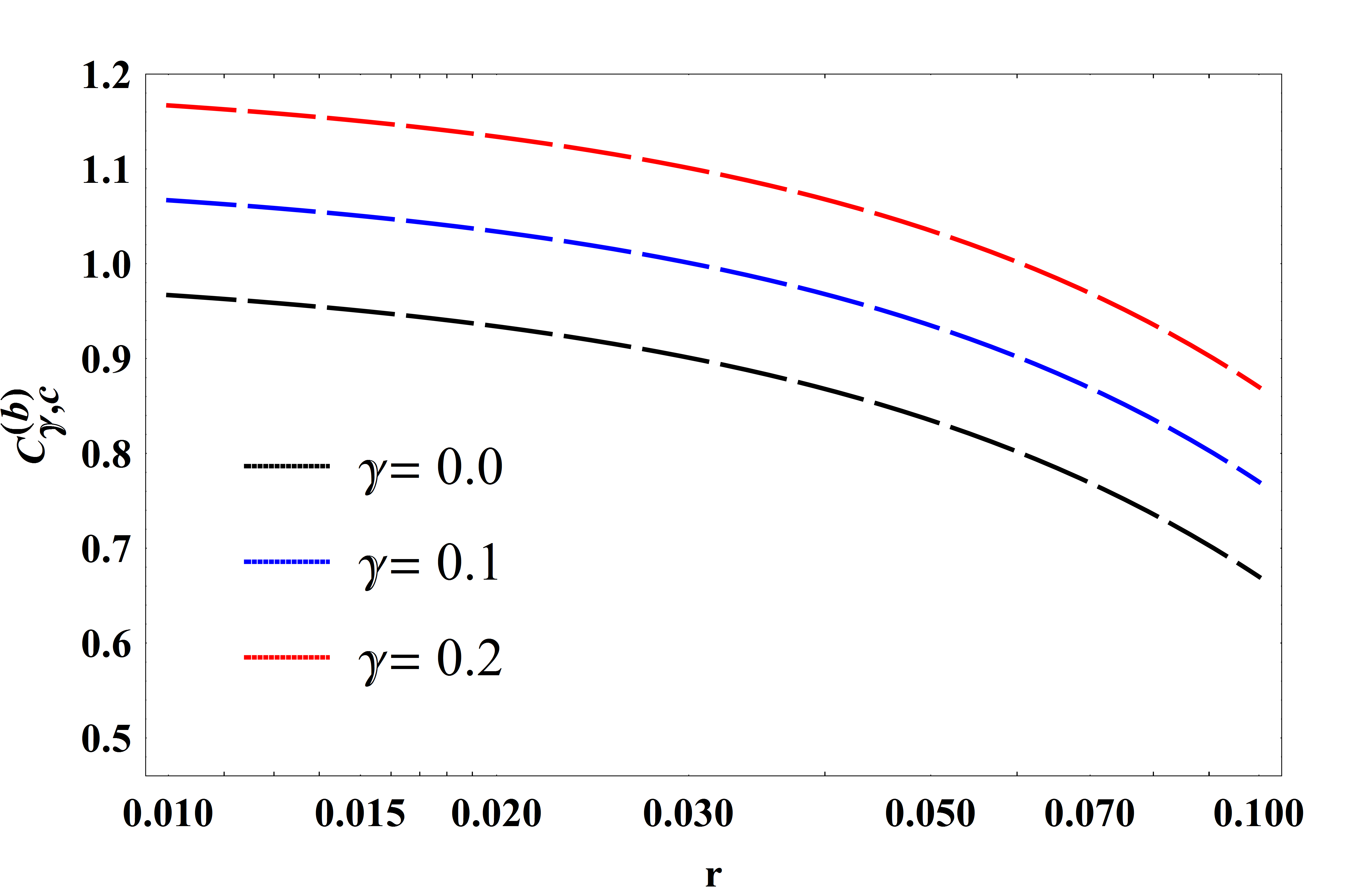}}\\

\end{tabular}
   \caption[]{Evolution of the correction term $C_{\gamma,c}^{(b)}$, versus the conformal anomaly coefficient $c$, the IG strength $\gamma$ and the tensor-to-scalar ratio $r$, respectively. Here we have imposed the constraint on the power spectrum of scalar perturbations from Planck data \cite{Akrami:2018odb}: $A_{s}^{2}=2.101 \times 10^{-9}$.}\label{figure3}
\end{figure*}

An upper bound on the conformal anomaly coefficient, $c_{max}$, is obtained by equating $U$ to unity and the standard cosmology is recovered for $c\ll c_{max}$ (i.e. $U\ll 1$). By using the latest Planck data \cite{Akrami:2018odb} ($A_{s}^{2}=2.101 \times 10^{-9}$) for $r=0.060$ and $\gamma=0.9$, we find that $c$ must satisfy the following condition
\begin{equation}\label{cmax}
  c\ll c_{max}= 1.92\times 10^{8}.
\end{equation}

In Figures \ref{fig1a}-\ref{fig1d}, we show the evolution of the dimensionless parameter $U$ Eq. \eqref{U} versus the conformal anomaly coefficient $c$ and the IG strength. We observe that the effect of the IG correction as well as the holographic cosmology starts from the values of the conformal anomaly coefficient $c>10^{7}$, otherwise the holographic cosmology and the IG correction have no effect on the standard cosmology dynamics. Therefore, we plot the evolution of the tensor-scalar ratio versus the dimensionless parameter $U$, Figs. \ref{fig1e} and \ref{fig1f}, in the range $c>10^{7}$. From these two figures we notice that the predicted values of the tensor-scalar ratio are included in the bound imposed by Planck data \cite{Akrami:2018odb}.\par
From now on we will consider the range $10^{7} <c<1.92\times 10^{8}$ for the conformal anomaly coefficient in order that holographic cosmology can leave its imprints on the spectrum of the gravitational waves.\par

The behavior of the correction terms $C_{\gamma,c}^{(a)}$ and $C_{\gamma,c}^{(b)}$ is shown in Fig. \ref{figure2} and Fig. \ref{figure3} respectively. From these figures we see that the effect of the holographic cosmology appears only for the values $c>10^{7}$ which confirms the aforementioned conclusion about the range of $c$. The maximal value of the conformal anomaly coefficient corresponding to $U=1$ is reflected by the maximum of the correction term $C_{\gamma,c}^{(a)}$ which is not defined (see Fig. \ref{fig2a}) and by the maximum of $C_{\gamma,c}^{(b)}$ which is equal to $0.5$ for $\gamma=0$ (see Fig. \ref{fig3a}). We notice also from Figs. \ref{figure2} and \ref{figure3} that for $0\leq\gamma<1$, we can always find a range of the conformal anomaly coefficient in which the effect of the IG is appreciable.\par

\subsection{Scalar inflation with quadratic potential}
\begin{figure*}
  \centering
 \begin{tabular}{ccc}
    \subfloat[ $r=0.06$ and $b=8\times10^{-6}$ \label{fig4a}]{\includegraphics[scale=0.5]{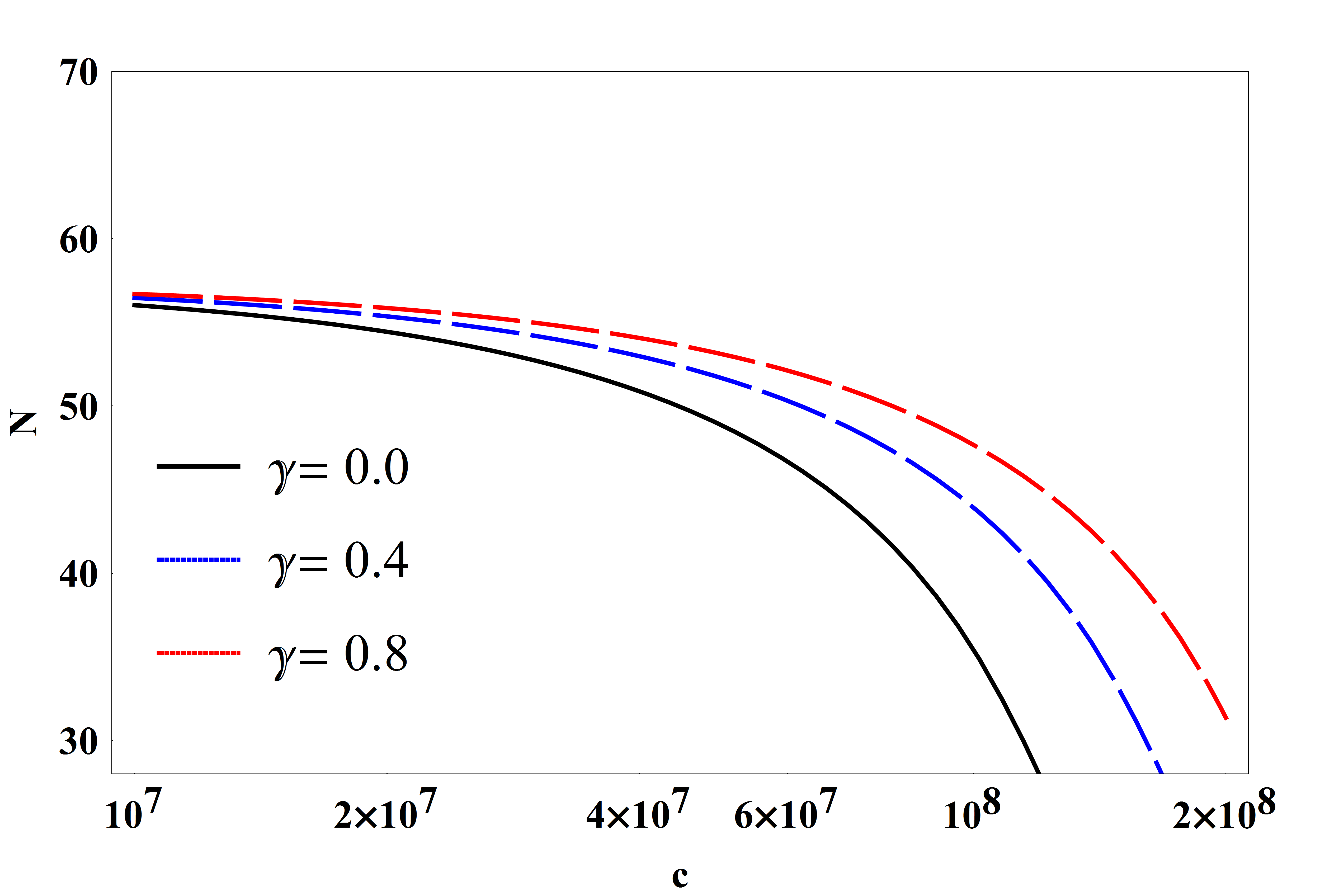}}&
     \subfloat[$\gamma=0.2$ and $b=7.8\times10^{-6}$\label{fig4b}]{\includegraphics[scale=0.5]{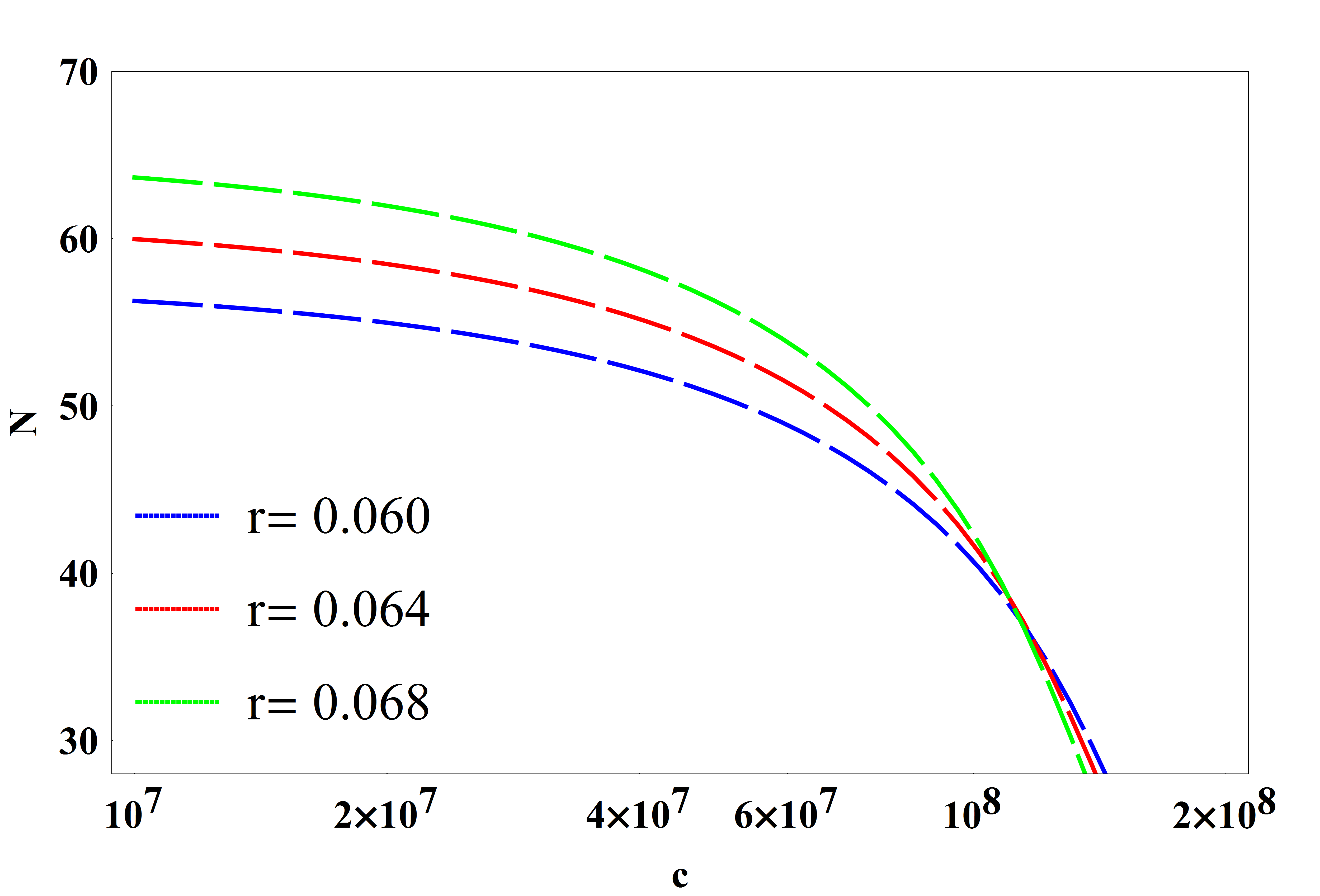}}&
      \subfloat[$\gamma=0.2$ and $r=0.06$\label{fig4c}]{\includegraphics[scale=0.5]{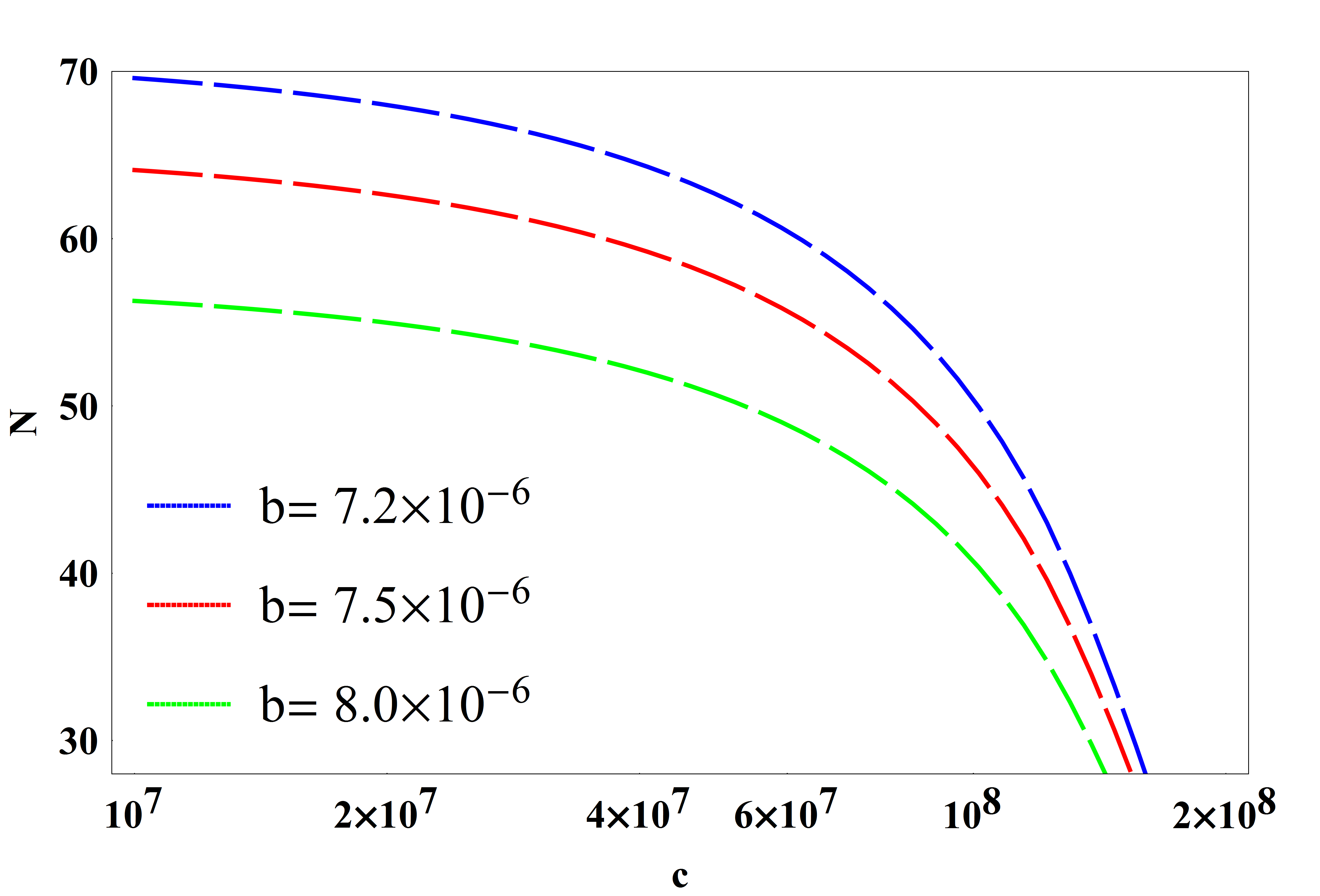}}\\
     \end{tabular}
   \caption[]{Evolution of  $N$ versus the conformal anomaly coefficient $c$.}\label{figure4}
\end{figure*}

In this subsection, we begin by specifying the functional form of the potential in order to calculate the spectrum of density perturbations and of gravitational waves, and then check their consistency with observations. Here we will consider the quadratic scalar potential given by the following expression
\begin{equation}\label{potential}
  V(\phi) = \frac{1}{2} \kappa_{4}^{-2} b^{2} \phi^{2},
\end{equation}
where $b$ is a dimensionless parameter. By substituting the quadratic potential Eq. \eqref{potential} in Eq. \eqref{efoldsslow}, the number of e-folds reads to
{\small
\begin{equation}\label{Nfinal}
  N=\frac{3(1+\gamma)}{4 cb^{2}}\left(\sqrt{1-U_{f}}-\sqrt{1-U_{i}}\\+\ln(\frac{1+\sqrt{1-U_{i}}}{1+\sqrt{1-U_{f}}})\right).
\end{equation}
}
By using equations \eqref{epsilon} and \eqref{potential} we rewrite the slow-roll parameter $\epsilon$ as
\begin{equation}\label{epsilonv}
  \epsilon = \frac{2cb^{2}}{3 (1+\gamma)} \frac{U}{\sqrt{1-U} (1-\sqrt{1-U})^{2}}.
\end{equation}
Inflation breaks down at $\epsilon =1$ and at the low energy limit we find $U_{f}=\frac{8 c b^{2}}{3}$, i.e. $\phi_{f}=  \frac{\sqrt{2}}{\kappa_{4}}$ which recovers the standard form of the scalar field for a quadratic potential at the end of inflation \cite{UrenaLopez:2007vz}.
 Fig. \ref{figure4} shows the evolution of the number of e-folds $N$ against the coefficient $c$. In Figs. \ref{fig4a}, \ref{fig4b}, \ref{fig4c}, the plot is for different values of the IG strength $\gamma$ ($r=0.06$, $b=8\times10^{-6}$), different values of the tensor to scalar ratio $r$ ($\gamma=0.2$, $b=8\times10^{-6}$) and for different values of the parameter $b$ ($\gamma=0.2$, $r=0.06$) respectively. We conclude from these figures that for $c< 5\times10^{7}$ the number of e-folds is in the range $50<N<70$ which is in good agreement with observations.\par

From Eq. \eqref{asmodi}, we derive the scalar perturbation at the crossing horizon as
\begin{equation}\label{Ascrossing}
  A_{s}^{2} = \frac{3 (1+\gamma) (1-\sqrt{1-U_{i}})^{3}}{400 c^{2} \pi^{2} b^{2} U_{i}}.
\end{equation}

The scalar spectral index Eq. \eqref{ns} at the crossing horizon is given by
\begin{equation}\label{nfinal}
  n_{s}=1-\frac{4cb^{2}}{3(\gamma+1)}\left(\frac{3+\sqrt{1-U_{i}}}{\sqrt{1-U_{i}}(1-\sqrt{1-U_{i}})}\right).
\end{equation}

\begin{figure*}
  \centering
    \begin{tabular}{ccc}
    \subfloat[\label{fig6a} ]{\includegraphics[scale=0.55]{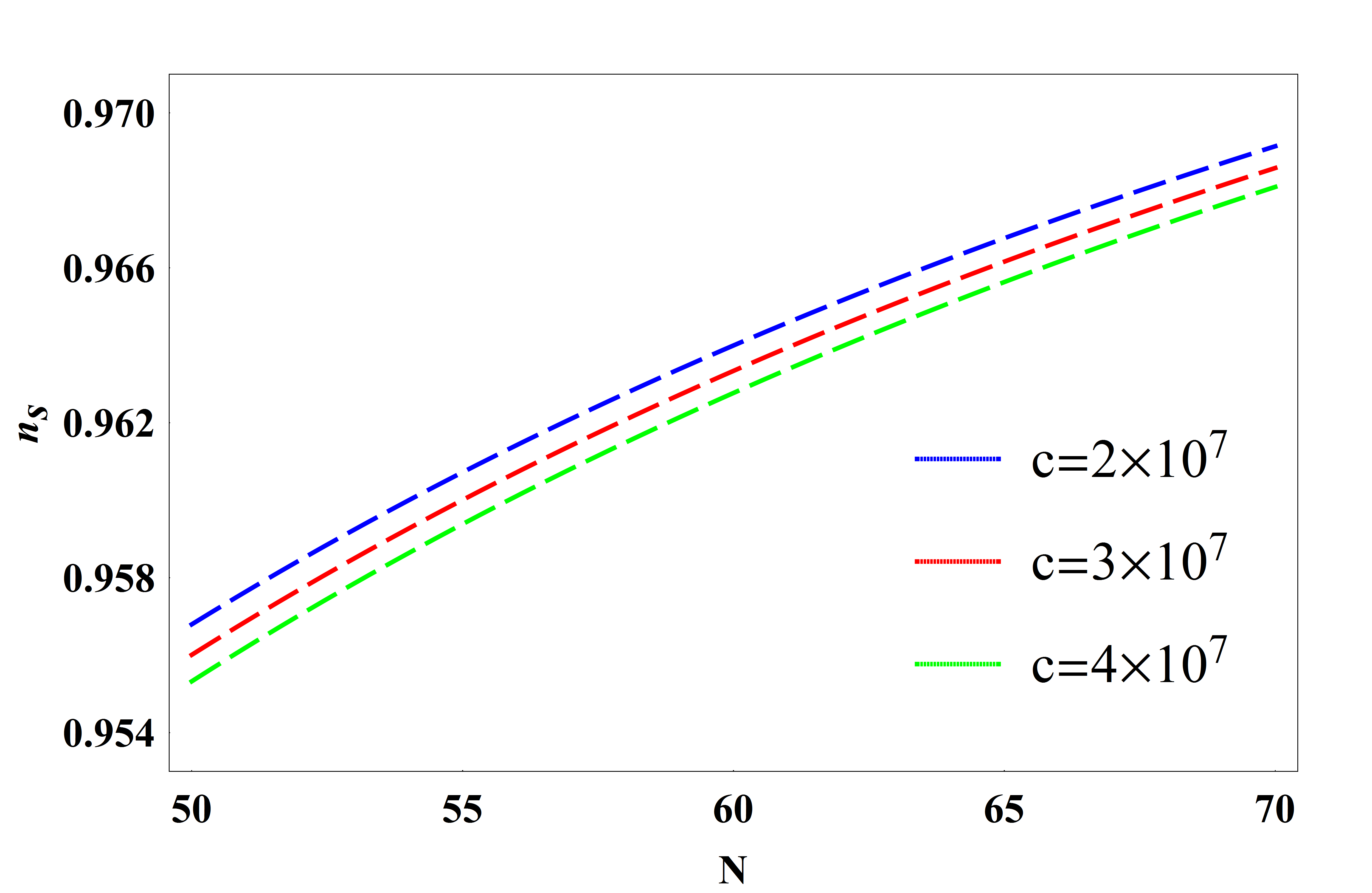}}&\qquad \qquad&
    \subfloat[\label{fig6b}]{\includegraphics[scale=0.55]{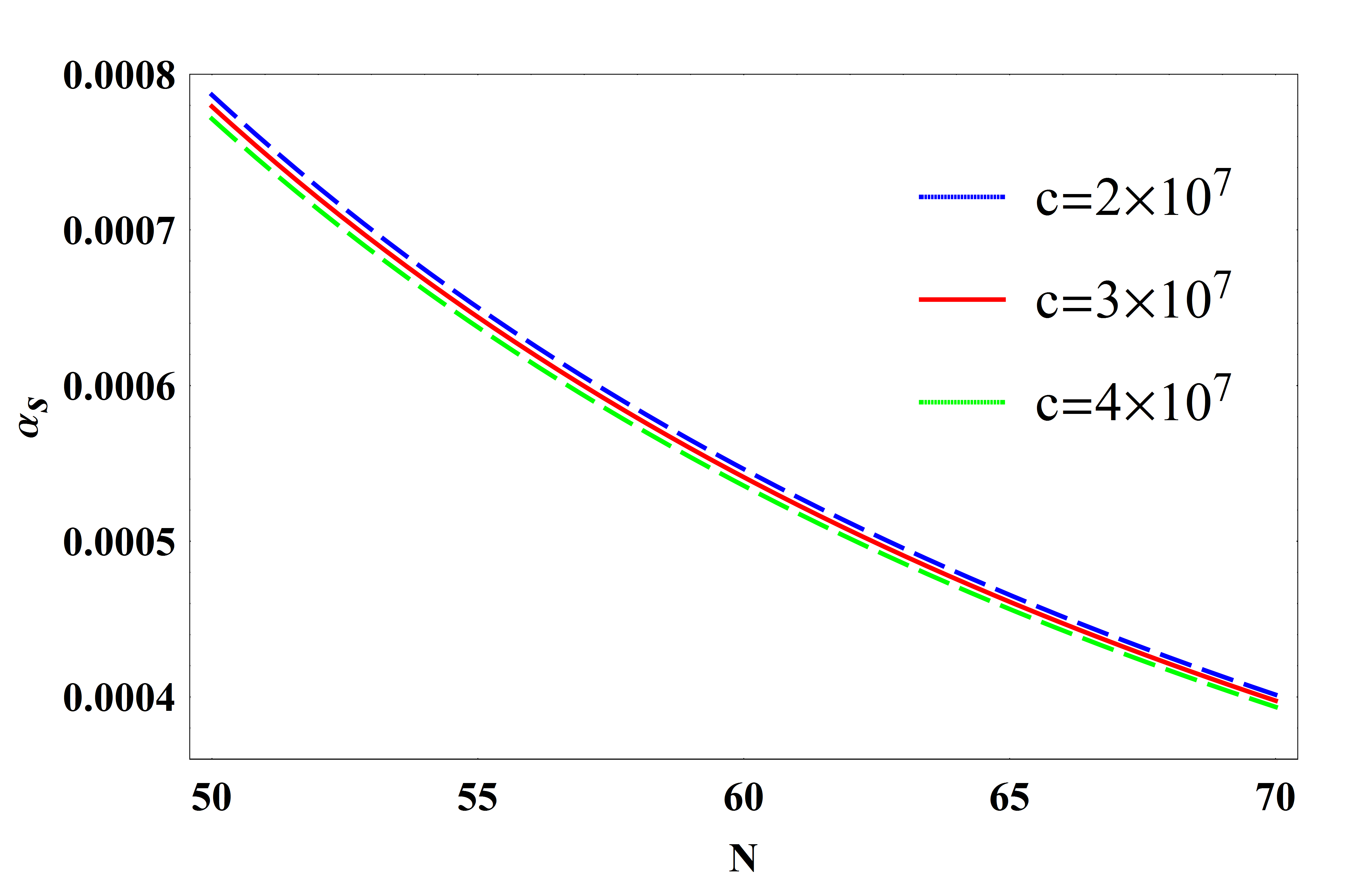}}\\
    \end{tabular}
   \caption[]{Evolution of  $n_{s}$ and $\alpha_{s}$ versus the number of e-folds $N$, for $\gamma=0.2$, $r=0.06$ and $b=10^{-6}$.}\label{figure5}
\end{figure*}

 Figure \ref{fig6a} shows the variation of the scalar spectral index versus the number of e-folds, for different values of the conformal anomaly coefficient $c$ for $\gamma=0.2$, $r=0.06$ and $b=10^{-6}$. The running of the spectral index $n_{s}$ defined as $\alpha_{s} = dn_{s}/dlnk$ can be expressed using Eq. \eqref{Nfinal} and Eq. \eqref{nfinal} as
\begin{widetext}
\begin{equation}\label{alphafinal}
  \alpha_{s}=-\frac{16 c^{2}b^{4}}{9 (1+\gamma)^{2}} \left[\frac{(1+\sqrt{1-U_{i}})  \left[-8 (1+\sqrt{1-U_{i}})+(12+8 \sqrt{1-U_{i}}-U_{i})U_{i}\right]}{(1-U_{i})^{\frac{3}{2}}U_{i}^{2}}\right].
\end{equation}
\end{widetext}

In Fig. \ref{fig6b} we show the evolution of the running $\alpha_{s}$ versus the e-folding number $N$, for different values of the conformal anomaly coefficient $c$ for $\gamma=0.2$, $r=0.06$ and $b=10^{-6}$. One can see from Figs. \ref{fig6a} and \ref{fig6b} that for $50<N<70$ the scalar spectral index $n_{s}$ and its running $\alpha_{s}$ are consistent with the Planck 2018 data \cite{Akrami:2018odb}.\par

In Table. \ref{table1}, we have summarized the predicted and the observed inflation parameters making use of Figs. \ref{fig6a} and \ref{fig6b}. These results are well supported by the Planck 2018 data.\par

\begin{figure*}
  \centering
  \begin{tabular}{ccc}
     \subfloat[ $\gamma=0.6$ and $b=10^{-6}$ \label{fig7a}]{\includegraphics[scale=0.65]{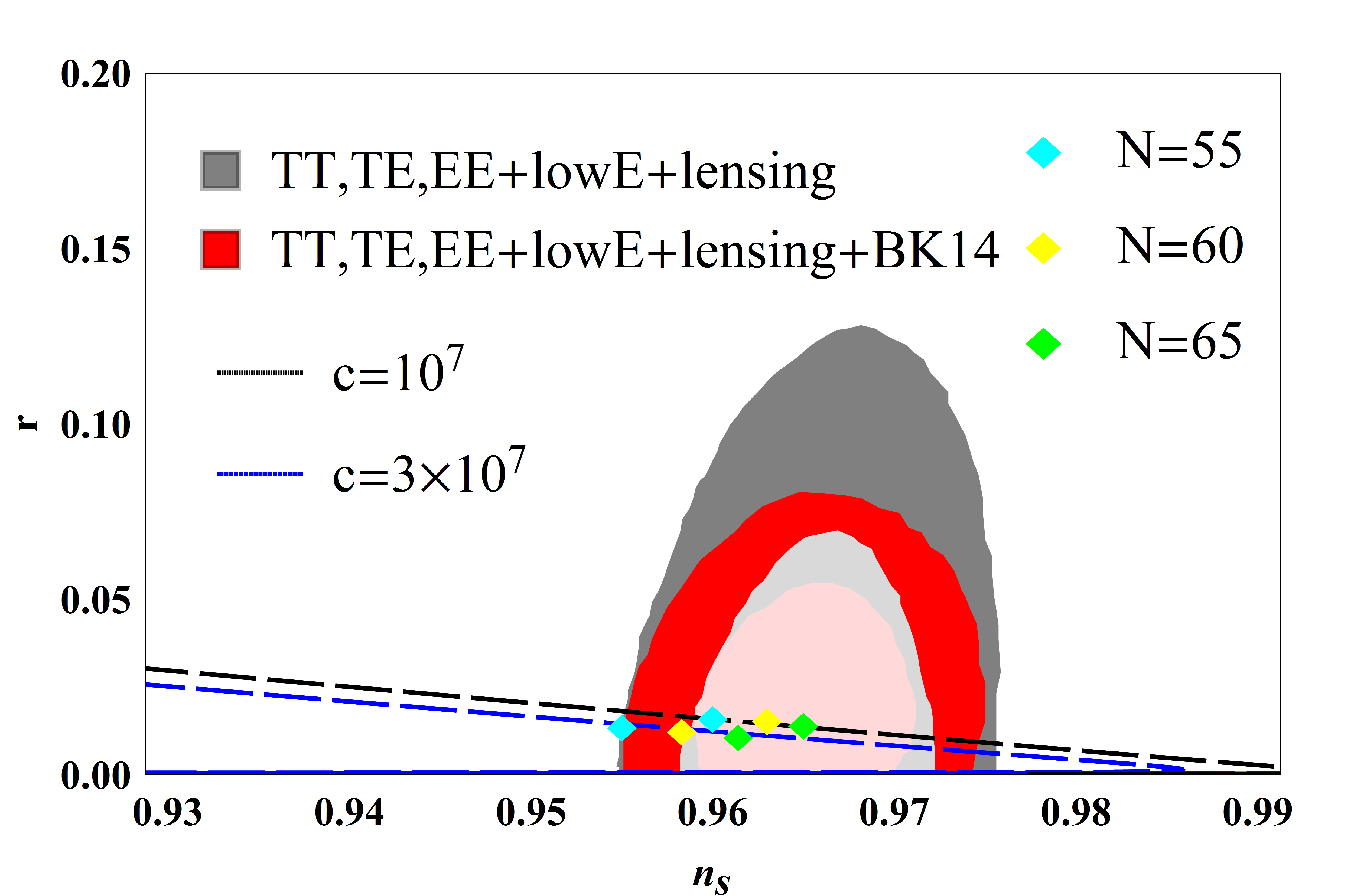}} &\qquad &
     \subfloat[$\gamma=0.6$ and $c=3\times10^{7}$ \label{fig7b}]{\includegraphics[scale=0.65]{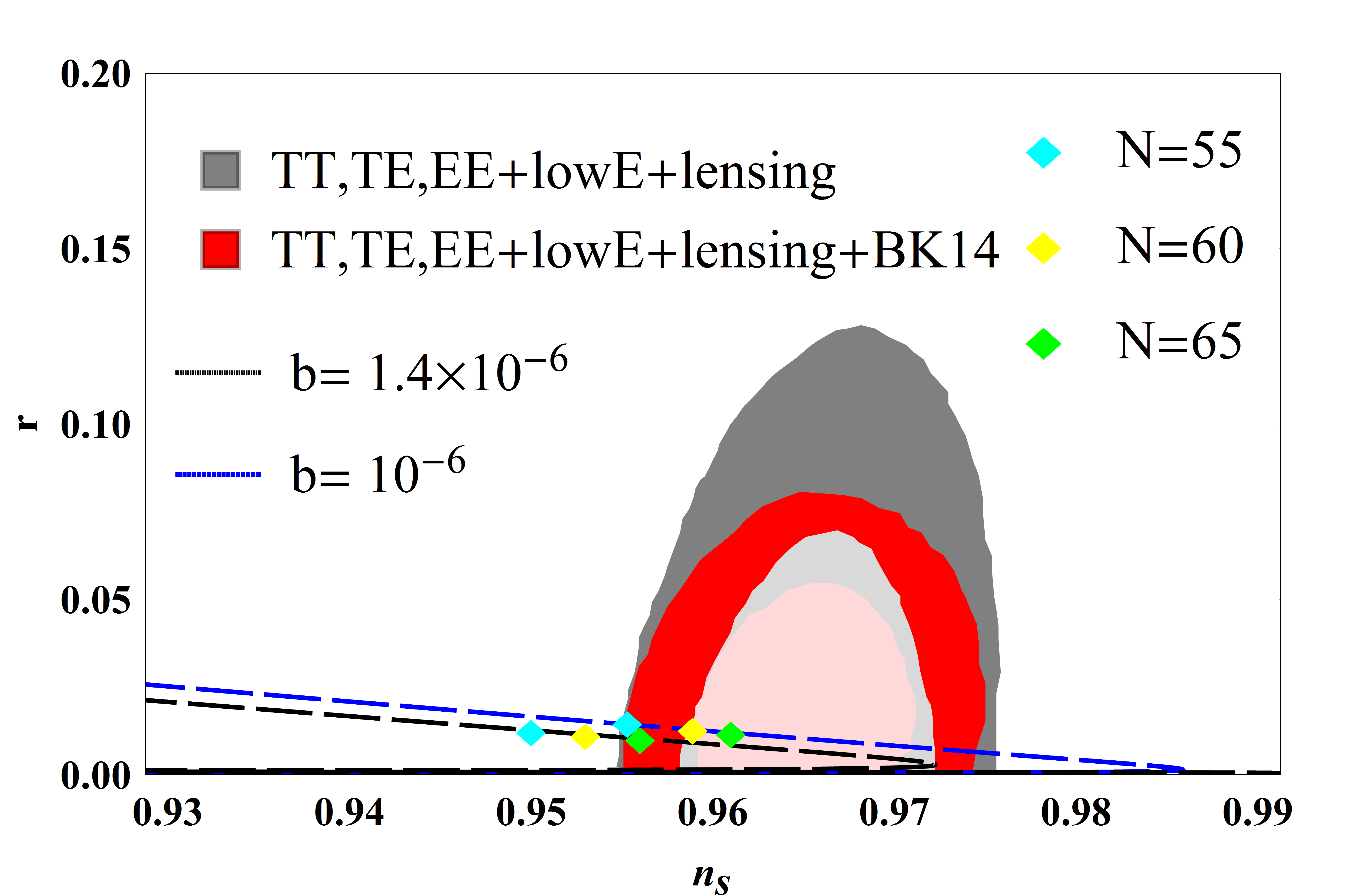}}\\
     \multicolumn{3}{c}{\subfloat[$c=3\times10^{7}$and $b=10^{-6}$ \label{fig7c}]{\includegraphics[scale=0.65]{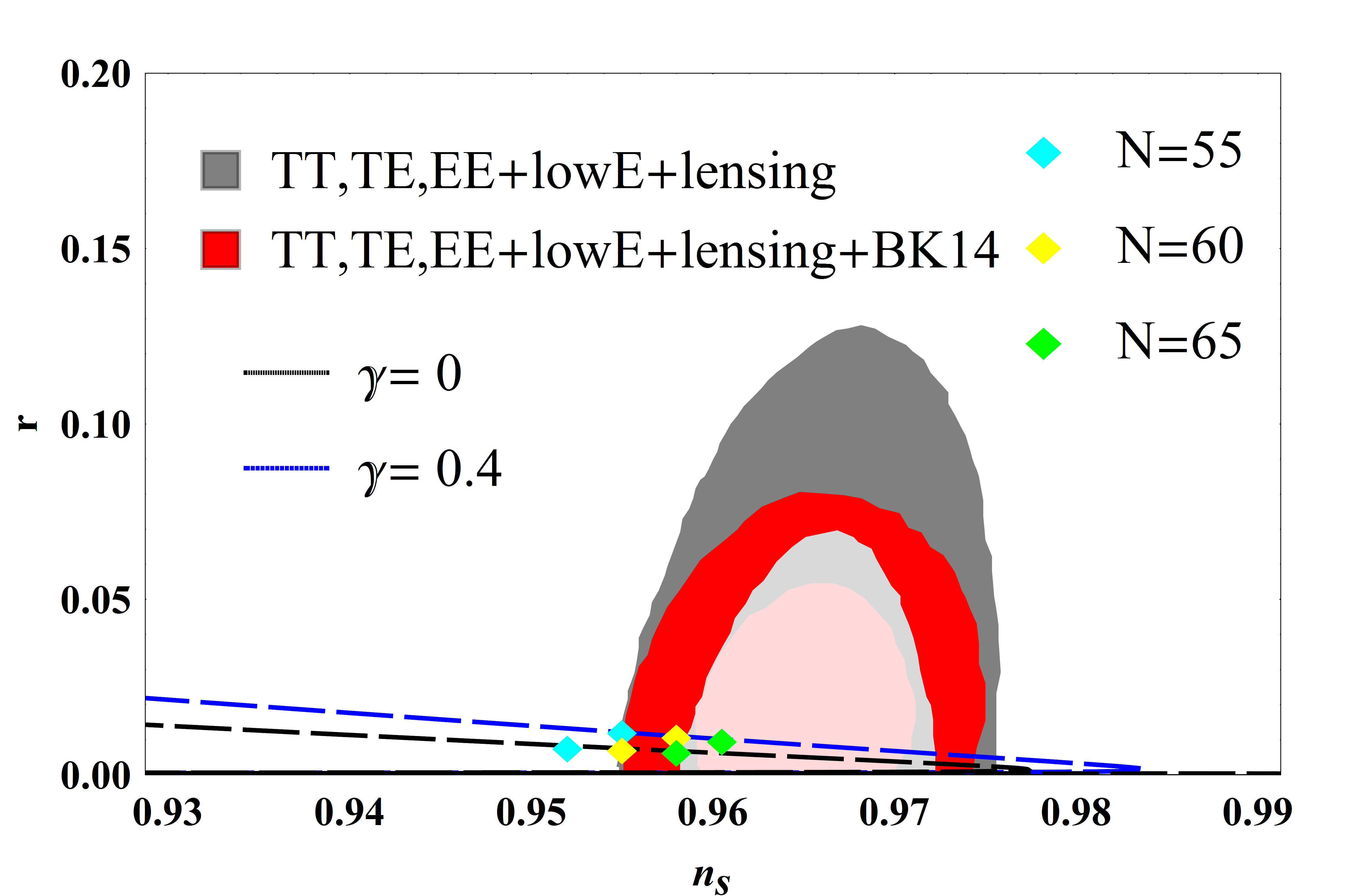}}}
\end{tabular}
   \caption[]{Plot of the tensor-to-scalar ratio $r$ against the scalar spectral index $n_{s}$. The marginalized joint $68$\% and $95$\% confidence
level contours $(n_{s}, r)$ using Planck alone and in combination with BK14 data.}\label{figure7}
\end{figure*}

Finally, the ratio between the amplitudes of tensor and scalar perturbations at the crossing horizon is given by
\begin{table*}
  \centering
\begin{tabular}{|c|c|c|c|c|c|}
  \hline
  & $N=55$  & $N=60$  & $N=65$  & $N=70$ & TT,TE,EE+lowE+lensing\\
   \hline
    $n_{s}$  & $0.9608$  & $0.9640$  & $0.9669$  & $0.9691$ & $ 0.9649\pm 0.0042$\\
\hline
 $\alpha_{s}$  & $0.00065$  & $0.00055$  & $0.00046$  & $0.0004$  & $-0.0045\pm 0.0067$\\
  \hline
\end{tabular}\caption{The values of some inflation parameters with a quadratic potential at the time that physical scales crossed the horizon for $b=1\times 10^{-6}$, c$=2\times 10^{7}$, $r=0.06$ and $\gamma=0.2$.}\label{table1}
\end{table*}
\begin{equation}\label{pj}
  r=\frac{16\pi cb^{2}}{3}\frac{(1+\sqrt{1-U_{i}})^{2}}{U_{i}}.
\end{equation}
Figures \ref{fig7a}, \ref{fig7b} and \ref{fig7c} show the $n_{s}-r$ contour plot for different values of $c$, $b$ and $\gamma$, in comparison with the observational data. The gray contour comes from the Planck TT, TE, EE, lowE+lensing data while the Planck TT, TE, EE, lowE+lensing+BK14 data are included in the red contour. The black and the blue dashed lines represent the theoretical predictions. One can see that the predicted parameters of the model lie within the $95\%$ C. L. region. In this figure we have highlighted three values of $N$ that are in the range $50<N<70$.  Moreover the parameter $b$ used in our plots lies in the range estimated by the authors in Ref. \cite{UrenaLopez:2007vz}.\par
\subsection{ Tachyonic inflation with an exponential potential}
In this subsection we adopt the exponential potential given by
\begin{equation}\label{po}
  V=V_{0}\exp(\frac{-\alpha}{\kappa_{4}}T),
\end{equation}
where $\alpha$ is a dimensionless parameter. This type of potential arises naturally from fundamental theories such as String theory/M theory \cite{tch}. With the aid of Eq. \eqref{po} and by using Eq. \eqref{efoldsslow} the number of e-folds for the tachyon field yields
{\small
\begin{equation}\label{t}
  N=\frac{3(\gamma+1)}{2 \alpha^{2}c}\left(\ln(\frac{1+\sqrt{1-U_{i}}}{1+\sqrt{1-U_{f}}})+\sqrt{1-U_{f}}-\sqrt{1-U_{i}}\right).
\end{equation}}
The slow roll parameters $\epsilon_{T}$ becomes
\begin{equation}\label{m}
  \epsilon_{T}=\frac{c\alpha^{2}}{3(1+\gamma)}\frac{U}{\sqrt{1-U}(1-\sqrt{1-U})^{2}},
\end{equation}
at the end of inflation ($\epsilon_{T}=1$), we find that at low energy, $U_{f}=4 c\alpha^{2}/3$. This gives $V_{f}=\alpha^{2}/2\kappa^{4}_{4}$ which is the same as the standard one \cite{Sami}.\par
\begin{figure*}
  \centering
 \begin{tabular}{ccc}
    \subfloat[$\alpha=1.1\times10^{-5}$ and $r=0.06$. \label{fig8a}]{\includegraphics[scale=0.5]{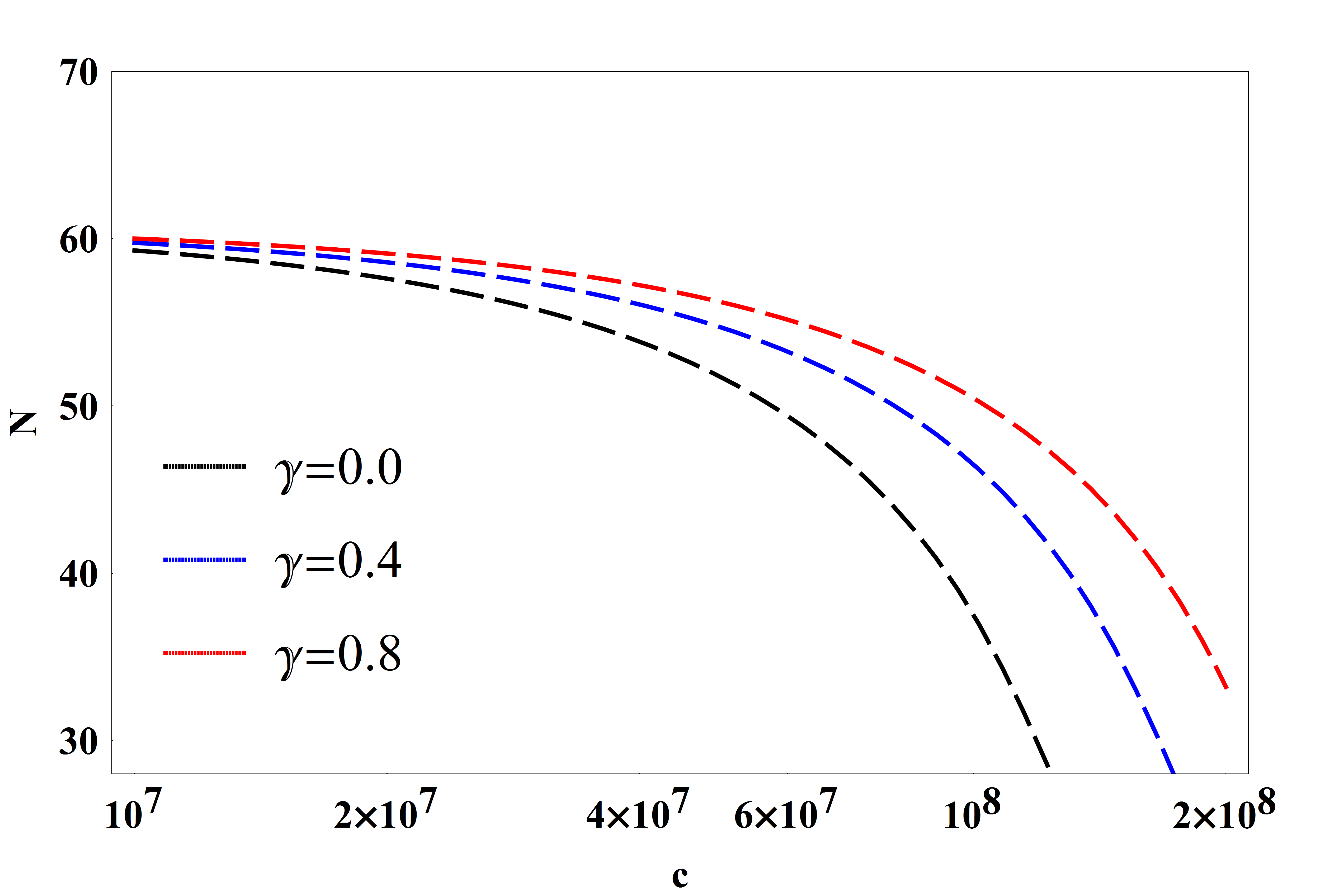}} &
     \subfloat[$\gamma=0.2$ and $r=0.06$.\label{fig8b}]{\includegraphics[scale=0.5]{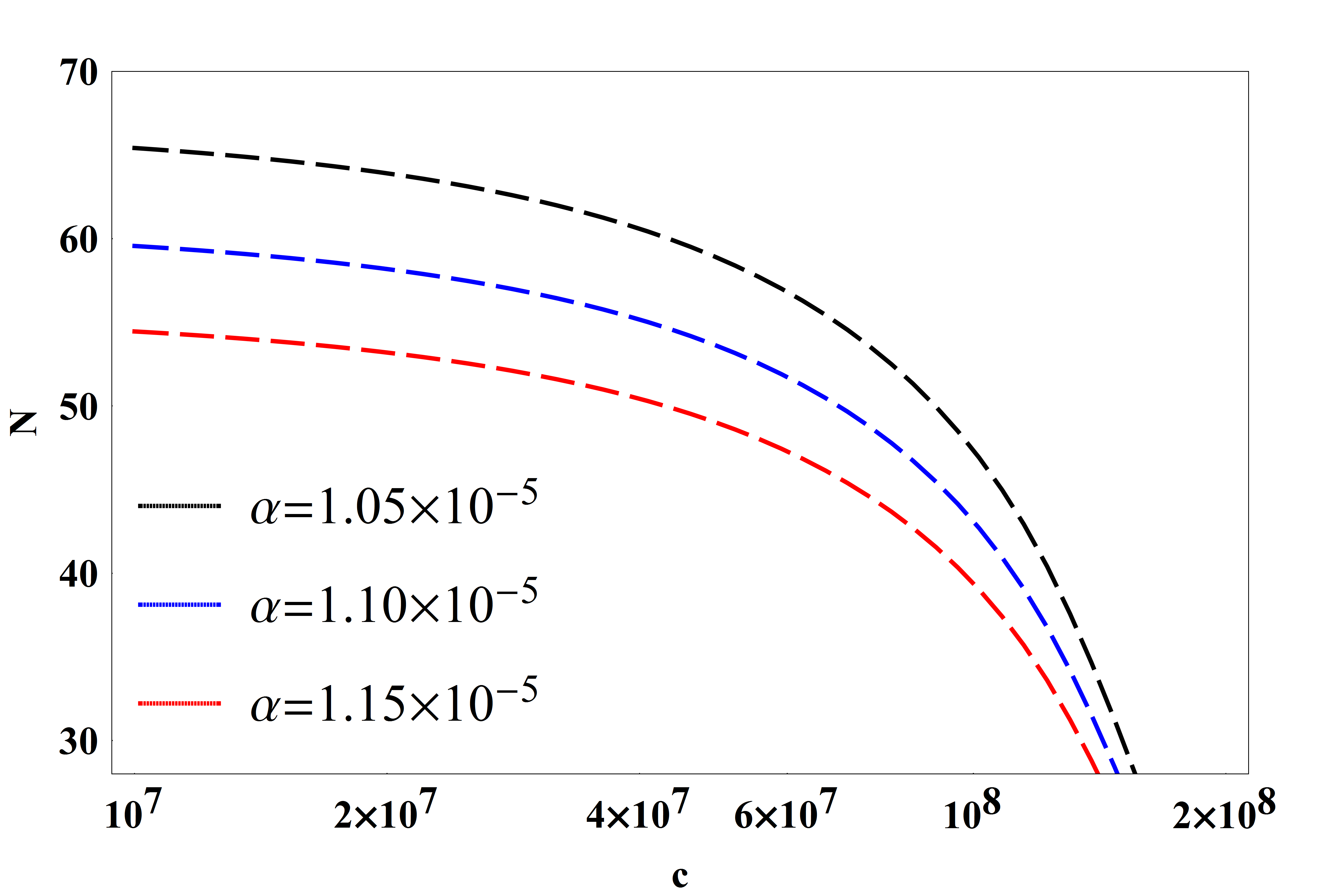}}&
      \subfloat[$\gamma=0.1$ and $\alpha=1.1\times10^{-5}$.\label{fig8c}]{\includegraphics[scale=0.5]{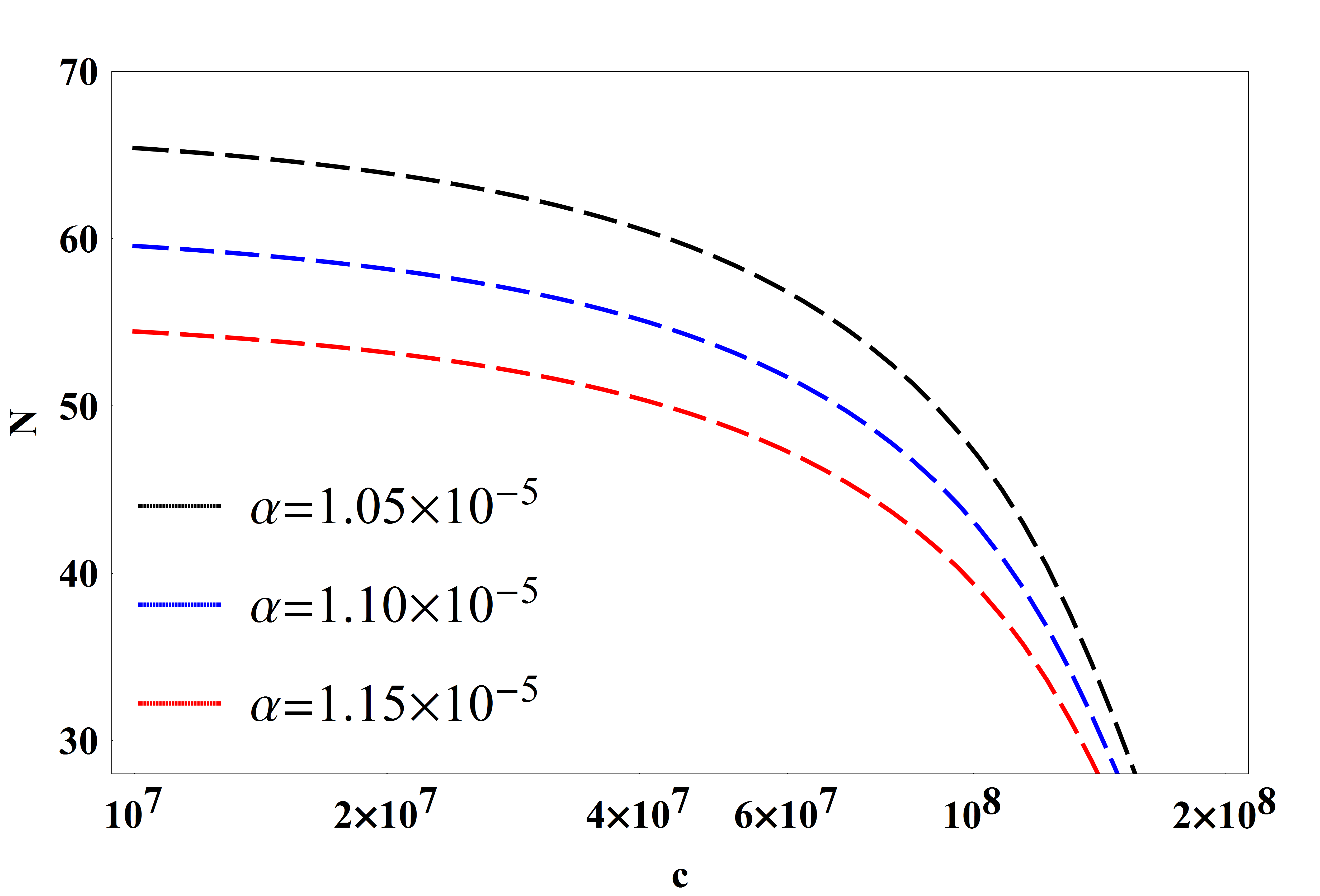}}
     \end{tabular}
   \caption[]{Evolution of  $N$ versus the conformal anomaly coefficient $c$.}\label{figure8}
\end{figure*}

Similarly, Fig. \ref{figure8} shows the variation of the e-folding number $N$ against the conformal anomaly coefficient for different values of $\gamma$, $\alpha$ and $r$. For $c< 5\times 10^{7}$ the number of e-folds $N$ lies well in the range $50<N<70$ favored by observational data.\par

From Eqs. \eqref{nsT} and \eqref{po} the scalar spectral index at the crossing horizon is given by
\begin{equation}\label{nss}
  n_{s}\simeq 1+\frac{2c \alpha^{2}}{3(1+\gamma)}\left(\frac{-3-\sqrt{1-U_{i}}}{\sqrt{1-U_{i}}(1-\sqrt{1-U_{i}})}\right).
\end{equation}
The running of the spectral index $n_{s}$ can be expressed using Eq. \eqref{nss} and Eq. \eqref{t} as
\begin{widetext}
\begin{equation}\label{alphafinal}
  \alpha_{s}=-\frac{4 c^{2}\alpha^{4}}{9 (1+\gamma)^{2}} \left[\frac{(1+\sqrt{1-U_{i}})  \left[-8 (1+\sqrt{1-U_{i}})+(12+8 \sqrt{1-U_{i}}-U_{i})U_{i}\right]}{(1-U_{i})^{\frac{3}{2}}U_{i}^{2}}\right].
\end{equation}
\end{widetext}
\begin{figure*}
  \centering
    \begin{tabular}{ccc}
    \subfloat[\label{fig9a}$n_{s}$ versus $N$ ]{\includegraphics[scale=0.55]{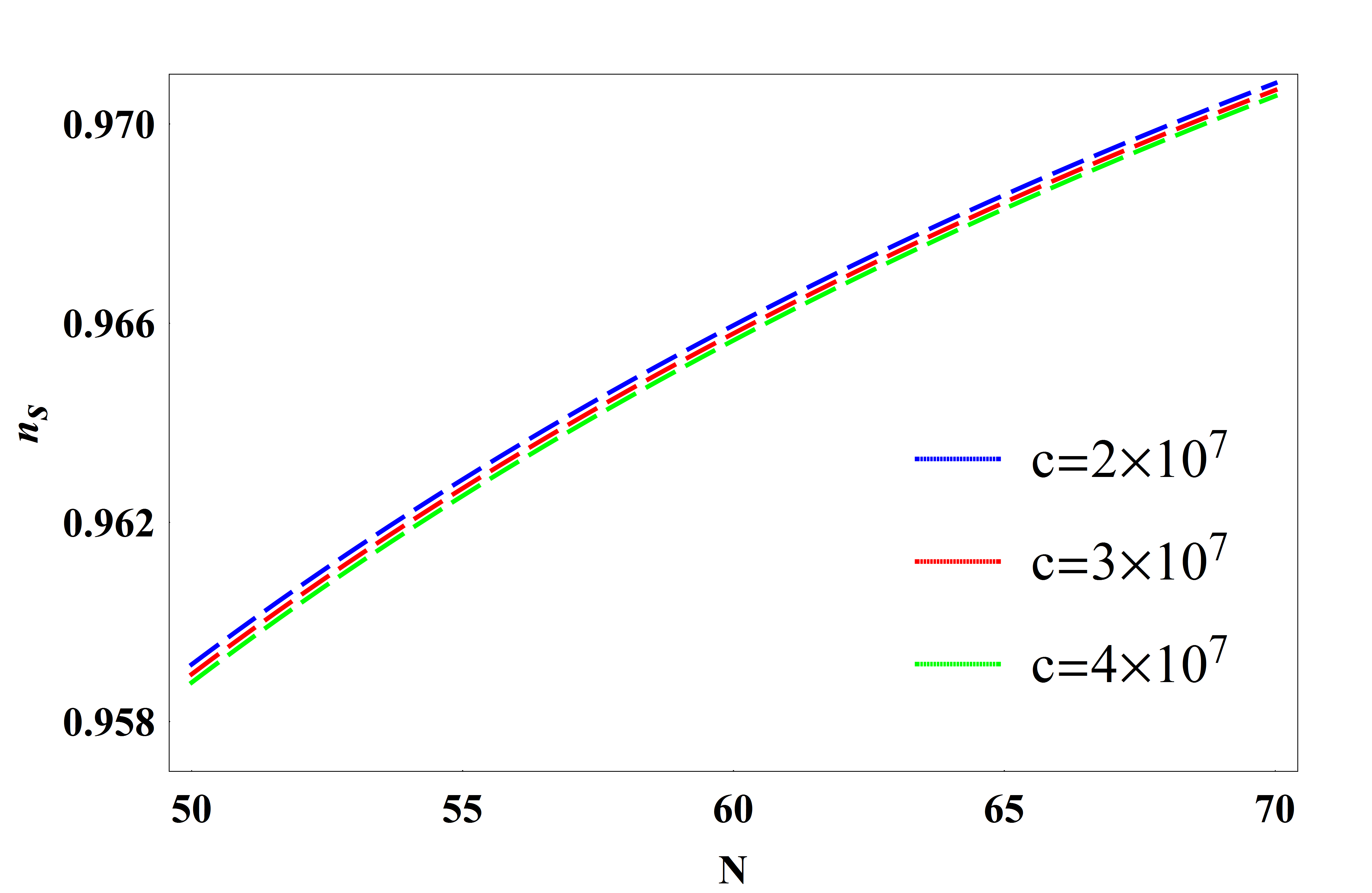}}&\qquad \qquad&
    \subfloat[\label{fig9b}$\alpha_{s}$ versus $N$ ]{\includegraphics[scale=0.55]{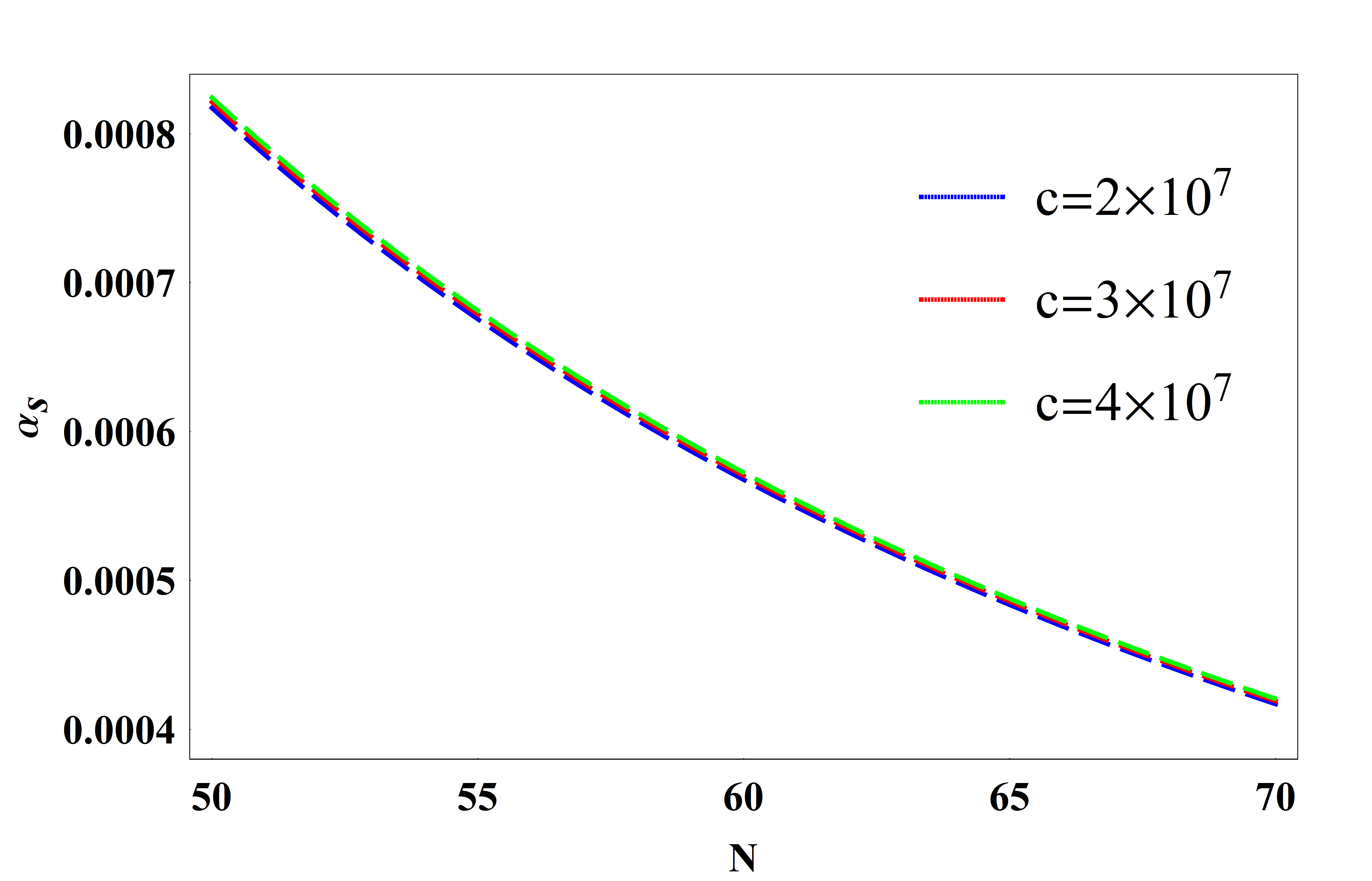}}\\
    \end{tabular}
   \caption[]{Evolution of  $n_{s}$ and $\alpha_{s}$ versus number of e-folds $N$, for $\gamma=0.2$, $r=0.06$ and $\alpha=10^{-6}$.}\label{figure9}
\end{figure*}
Figures \ref{fig9a} and \ref{fig9a} show, respectively,  the behavior of the spectral index $n_{s}$ and its running with respect to the number of e-folds $N$ for different values of $c$ and for $r=0.06$, $\gamma=0.2$ and $\alpha=10^{-6}$. While the parameter $n_{s}$ increases, its running decreases with the increment of the number of e-folds. Fig. \ref{figure9} shows the good agreement with the observationally viable values of the scalar spectral index and its running based on Planck 2018 data. Table \ref{table2} summarizes the predicted and the observed values of the scalar spectral index and its running. In this regard, we notice a good agreement between our predicted model parameters and the Planck 2018 data.
\begin{table*}
  \centering
\begin{tabular}{|c|c|c|c|c|c|}
  \hline
  & $N=55$  & $N=60$  & $N=65$  & $N=70$ &TT,TE,EE+LowE+Lensing+BAO\\
   \hline
    $n_{s}$  & $0.9609$  & $0.9643$  & $0.9668$ & $0.9691$ & $ 0.9649\pm 0.0042$ \\
     \hline
    $\alpha_{s}$  & $0.00067$  &  $0.00056$ & $0.00048$  & $0.00041$ & $-0.0041\pm 0.0067$ \\
  \hline
\end{tabular}\caption{The values of some inflation parameters with an exponential potential at the time that physical scales crossed the horizon for $\alpha= 10^{-6}$, $c=2\times 10^{7}$, $r=0.06$ and $\gamma=0.2$.}\label{table2}
\end{table*}

Finally, the ratio between the amplitude of tensor and scalar perturbations at the crossing horizon is given by
\begin{equation}\label{pj}
  r=\frac{32 \pi c\alpha^{2}}{75}\frac{(1+\sqrt{1-U_{i}})^{2}}{U_{i}}.
\end{equation}
\begin{figure*}
  \centering
  \begin{tabular}{ccc}
    \subfloat[$\alpha=10^{-6}$ and $\gamma=0.6$ \label{fig11a}]{\includegraphics[scale=0.65]{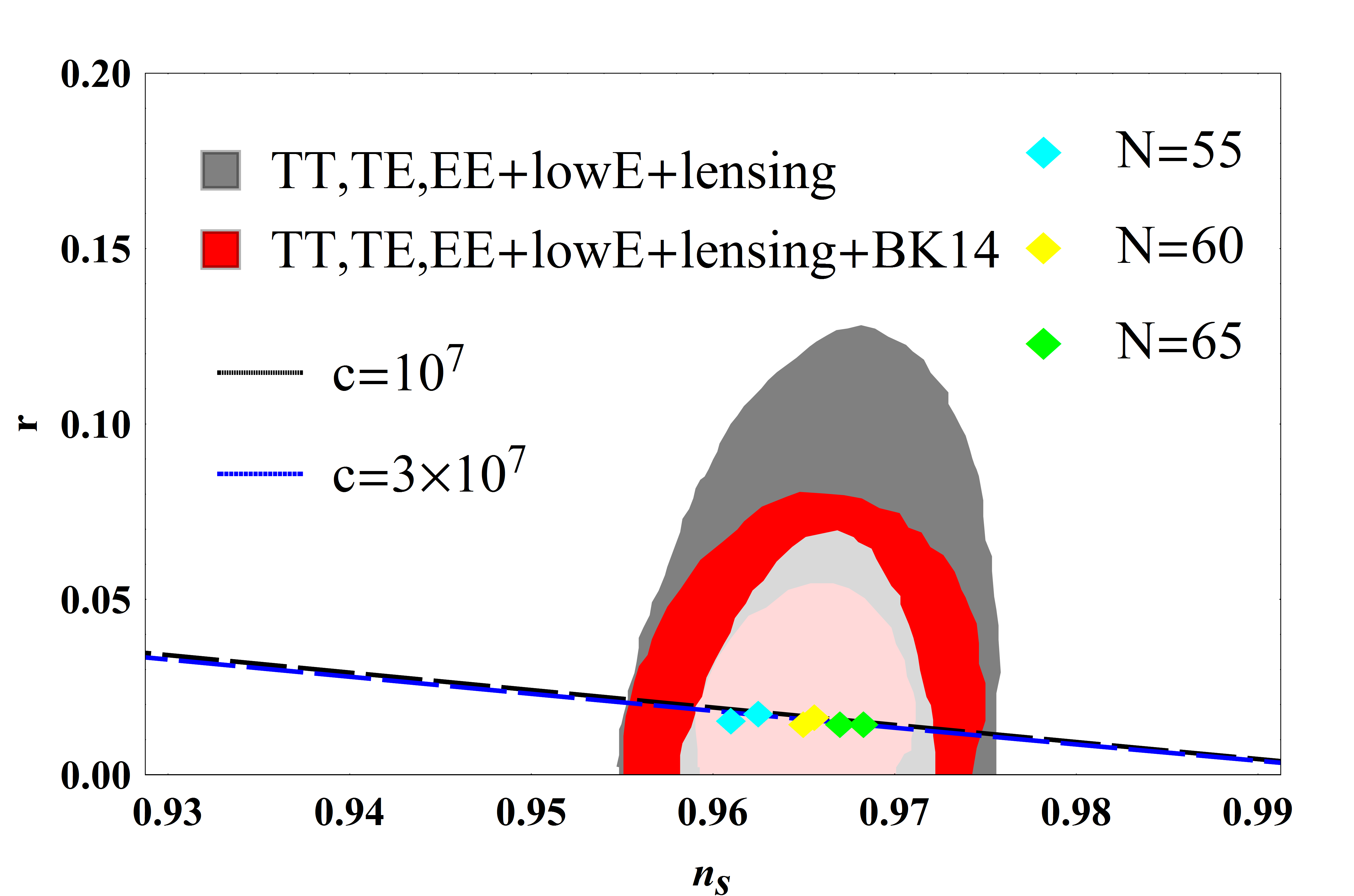}} & \qquad &
     \subfloat[$\gamma=0.6$ and $c=3\times10^{7}$\label{fig11b}]{\includegraphics[scale=0.65]{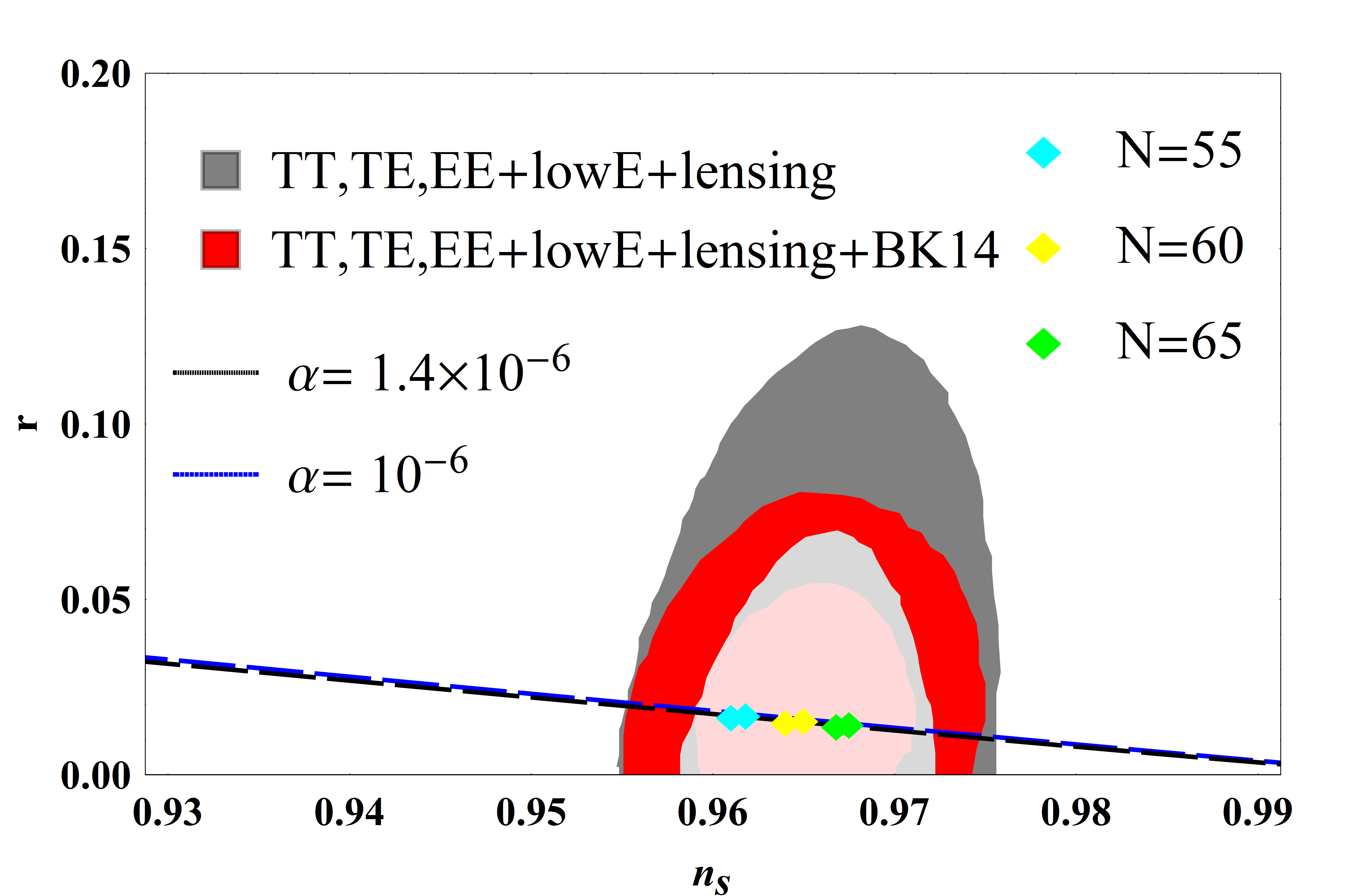}}\\
\multicolumn{3}{c}{\subfloat[$c=3\times10^{7}$and $\alpha=10^{-6}$ \label{fig11c}]{\includegraphics[scale=0.65]{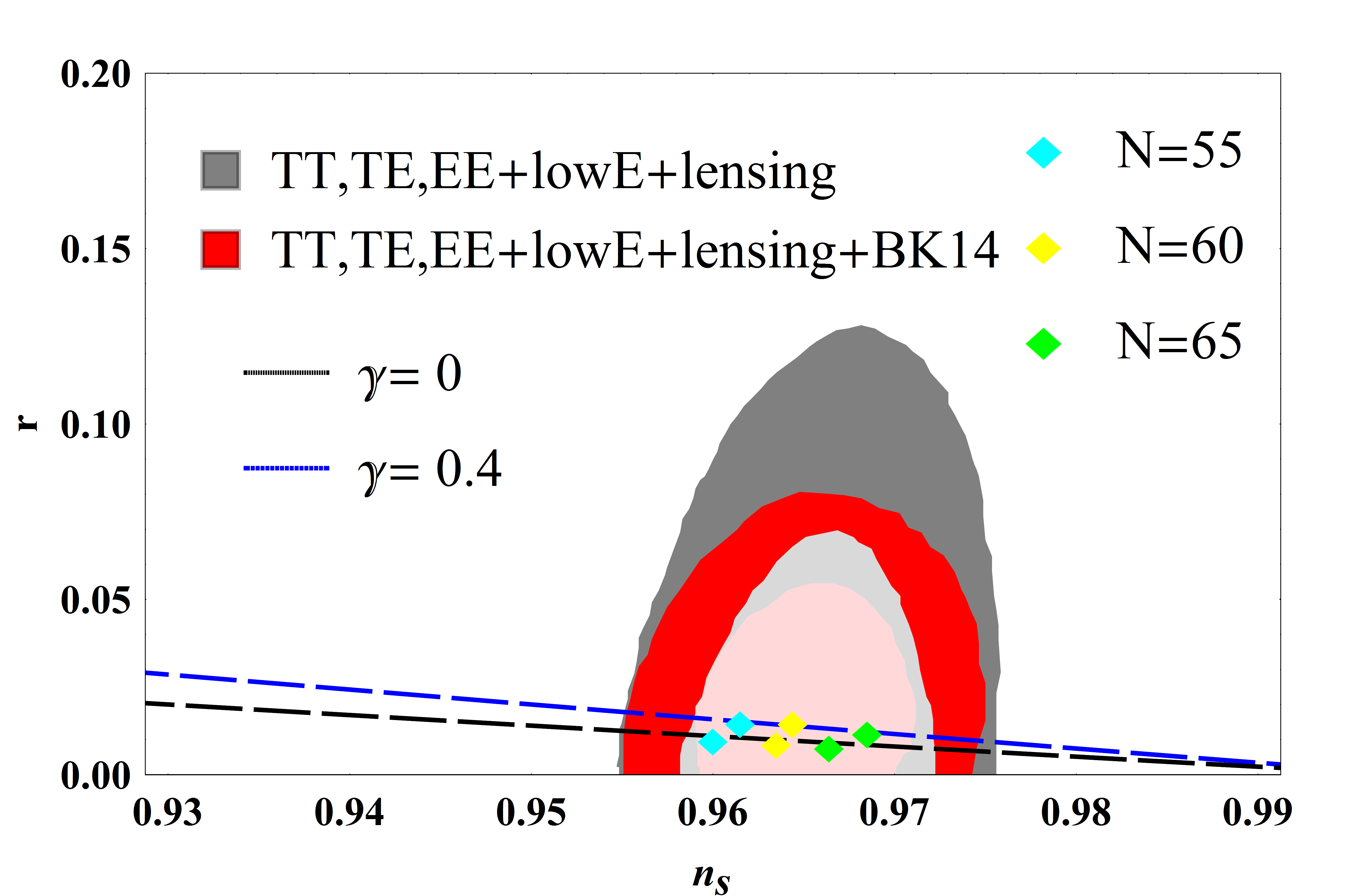}}}
\end{tabular}
   \caption[]{Plot of the tensor-to-scalar ratio $r$ against the scalar spectral index $n_{s}$. The marginalized joint $68$\% and $95$\% confidence
level contours $(n_{s}, r)$ using Planck alone and in combination with BK14 data.}\label{figure11}
\end{figure*}
\begin{figure}
{\includegraphics[scale=0.65]{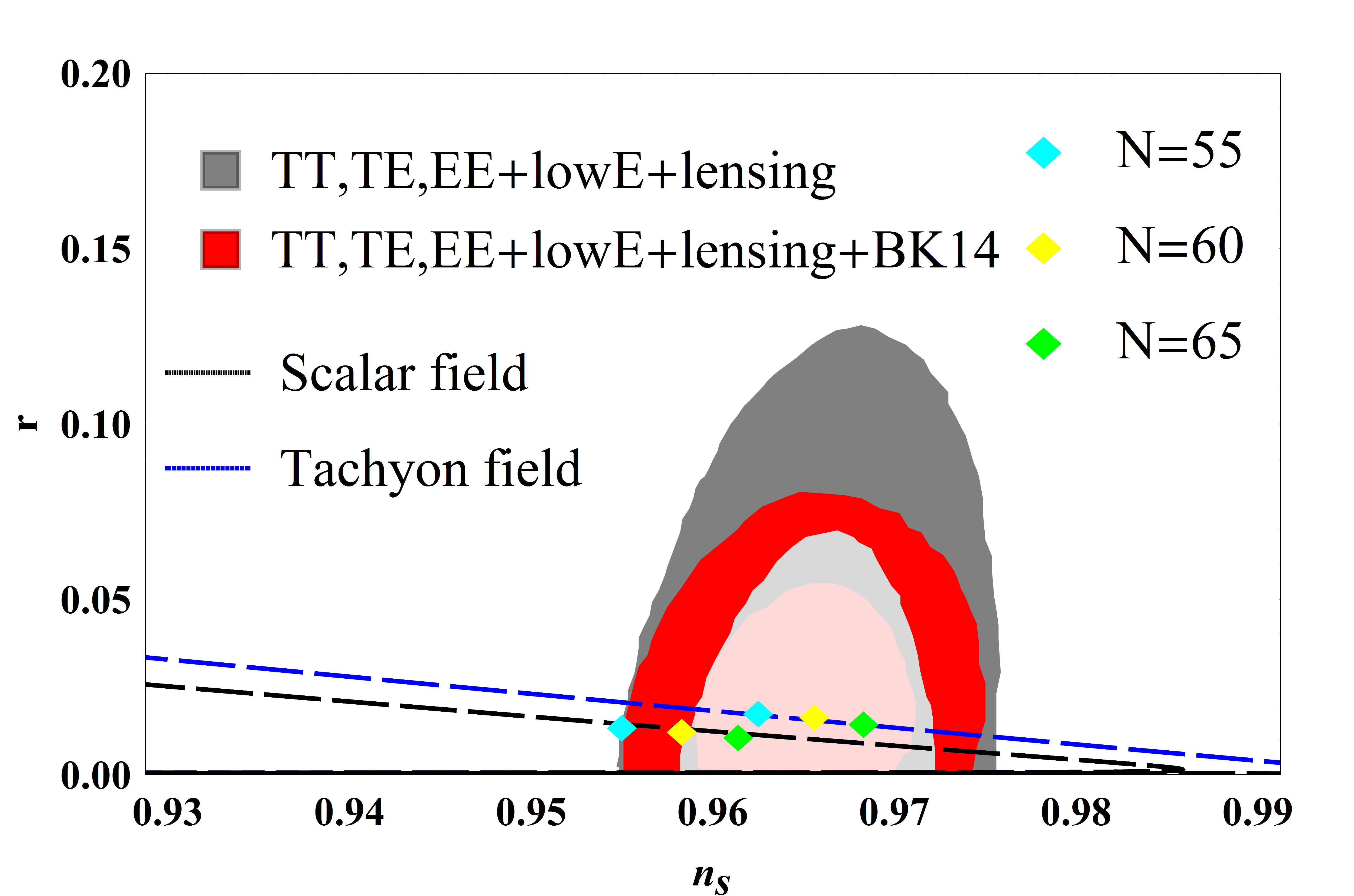}}
\caption[]{Plot of the tensor-to-scalar ratio $r$ against the scalar spectral index $n_{s}$ for both tachyon and scalar fields, for $c=3\times10^{7}$, $\alpha=b=10^{-6}$ and $\gamma=0.6$. The marginalized joint $68$\% and $95$\% confidence level contours $(n_{s}, r)$ using Planck alone and in combination with BK14 data.}\label{figure12}
\end{figure}

Figure \ref{figure11} shows Constraints from the Planck TT, TE, EE+lowE+lensing data (gray contour) and Planck TT, TE, EE+lowE+lensing+BK14 data (red contour) in the $n_{s}-r$ plane. Figs. \ref{fig11a}, \ref{fig11b} and \ref{fig11c} are plotted for different values of $c$, $\alpha$ and $\gamma$ respectively. The black and blue dashed lines represent theoretical predictions of the model parameters. We can see that our predicted parameters lie inside the 95\% C. L. of the Planck data. In this figure we have highlighted three values of $N$ that are in the range $50<N<70$. Furthermore, the value of the parameter $\alpha$ used in these plots lies in the range estimated by authors in Ref. \cite{Bouabdallaoui:2016izz}.\par
To compare the consistency between theoretical predictions and observations of the tachyon and the scalar fields, we plot the $n_{s}-r$ plane in Fig. \ref{figure12}. From this figure, one can notice a better agreement, at $95$\% C.L., for the tachyon than the scalar field for the selected numbers of e-folds. Furthermore, the scalar field shows smaller values of the tensor to scalar ratio compared to the tachyon field But both of them  still fit the data quit well.\par
\section{Summary and conclusions}\label{sec5}
In this paper, we have studied the cosmological inflation on a de Sitter brane with an induced gravity correction within the holographic cosmology. By considering a universe filled by both scalar and tachyon fields separately, we expressed the field equations in the slow-roll approximation. We have assumed a quadratic potential, for the scalar field, which has proven to be quite successful in constraining model's parameters, then we have adopted an exponential potential for the tachyon field.\par

We have found that, the effect of the IG and the holographic cosmology is not negligible in the range of $c>10^{7}$ as can be seen from Fig. \ref{figure1}. We have also noted that the main perturbation parameters such as the scalar spectral index, its running and the tensor-to-scalar ratio are modified by some correction terms which are illustrated in Figs. \ref{figure2} and \ref{figure3}. As these figures show, our results are affected by induced gravity corrections for $0<\gamma<1$ in addition to the holographic cosmology for $c> 10^{7}$.\par

In order to check the consistency of the predicted parameters with observation, we have compared our results against those of Planck 2018 data by plotting the Planck confidence contours in the plane of $n_{s}-r$. In this regard, the comparison indicates that the predicted parameters are consistent with the observational data, in the appropriate range of the conformal anomaly coefficient and the IG term, for both scalar and tachyon fields Figs. (\ref{figure7}-\ref{figure11}).\par

We have noticed a better agreement to the observational data, at $95$\% C.L., for the tachyon than the scalar field for the selected numbers of e-folds. Furthermore, while the scalar field shows smaller values of the tensor to scalar ratio compared to the tachyon field, both of the tachyon and the scalar fields are in good agreement with the Planck 2018 data. On the other hand, it can be also noted that the presence of the IG correction allows to expand the range of the conformal anomaly coefficient, see Eq.\eqref{cmax}, compared to the range found by authors in Ref. \cite{Bouabdallaoui:2016izz}.\par

\section*{Acknowledgment }
The authors would like to thank Mariam Bouhmadi-L\'{o}pez for many useful discussions and suggestions.\\

\end{document}